\documentclass[lettersize,journal]{IEEEtran}
\usepackage{cite}             
\usepackage{ifthen}
\newcounter{twoColumn}
\usepackage[cmex10]{amsmath}  
\usepackage{mleftright}       
\mleftright                   
\usepackage{booktabs}         

\usepackage[utf8]{inputenc}
\usepackage[fleqn]{nccmath} 
\usepackage{amssymb} 
\usepackage{amsmath}
\usepackage{dsfont} 
\usepackage[dvipsnames]{xcolor}
\usepackage{amsthm} 
\usepackage{aligned-overset}
\usepackage[bookmarks=false]{hyperref}
\usepackage{algpseudocode}
\usepackage{algorithm}

\newtheoremstyle{mytheorem}
  {0pt}
  {0pt}
  {\itshape}
  {}
  {\bfseries}
  {:}
  {0.5em}
  {}

\theoremstyle{mytheorem}

\newtheorem{theorem}{Theorem}
\newtheorem{lemma}{Lemma}
\newtheorem{corollary}{Corollary}

\newtheorem{remark}{Remark}
\newtheorem{definition}{Definition}

\newcommand{\abs}[1]{\left|#1\right|}

\NewDocumentCommand\sqn{mg}{%
    \|\mathbf{#1}_{\IfNoValueTF{#2}{}{#2}}\|^2%
}

\usepackage{subcaption}
\usepackage{graphicx}
\graphicspath{{figs/}}
\usepackage{tikz}
\begin{document}

\title{Perfectly Covert Communication Assisted by an Intelligent Reflecting Surface}

%
\author{ 
\IEEEauthorblockN{Or Elimelech and Asaf Cohen} 
\IEEEauthorblockA{
\\Ben-Gurion University of the Negev}
}
\maketitle

\thispagestyle{plain} 
\pagestyle{plain}    
%
\begin{abstract}
\textcolor{black}{
This work investigates perfectly covert communication assisted by a passive Intelligent Reflecting Surface (IRS).
In contrast to most existing IRS-assisted covert communication studies, which allow a nonzero detection leakage and optimize an epsilon-covertness constraint, we study the stricter regime in which the received signal component at the warden is completely canceled.
We first derive a necessary and sufficient condition for perfect covertness and characterize its feasibility under Rayleigh fading.
For the case of two reflecting elements, we provide a closed-form characterization of all feasible IRS phase configurations.
For a general number of reflecting elements, we prove that the perfect-covertness condition is eventually satisfied almost surely as the number of IRS elements grows.
To construct such configurations, we distinguish between the full Bob-aware design problem and the perfect-covertness feasibility subproblem, and formulate the latter as a warden-signal nulling problem. 
We then propose a gradient-based IRS phase-design algorithm with per-iteration computational complexity $O(N)$ and prove that, with random initialization, it converges to a global minimizer with probability one over the initialization set.
The numerical results show that Bob-aware initialization preserves the legitimate link while driving Willie leakage to the numerical floor, and further evaluate multi-antenna Willie and imperfect-CSI settings.
Finally, to address practical limitations such as imperfect channel state information and finite detector resolution, we introduce operational perfect covertness and derive a robust transmit-power condition that guarantees indistinguishability at the warden under bounded CSI uncertainty.}
\end{abstract}
\section{Introduction}
Covert communication, also known as low probability of detection (LPD) communication, enables two legitimate parties, Alice and Bob, to exchange information while effectively hiding the existence of their transmission from a vigilant warden, Willie. 
The fundamental limits of such communication were established in the seminal works on the square-root law \cite{bash2012square,bash2013limits,bash2015hiding}, which show that for an AWGN channel with a warden and a standard covert constraint, Alice can reliably and covertly transmit at most $\mathcal{O}(\sqrt{n})$ bits in $n$ channel uses, implying a vanishing rate as $n \to \infty$.

To overcome this zero-rate limitation, subsequent research focused on scenarios in which the legitimate parties have an advantage over the warden.
It has been shown that a positive (non-vanishing) covert rate is achievable in the presence of noise uncertainty \cite{he2017covert, goeckel2015covert}, lack of synchronization \cite{bash2014lpd}, or channel uncertainty \cite{lee2014achieving, che2014reliable}.
Additionally, external sources of uncertainty, such as jamming or interference, can facilitate covertness.
For instance, \cite{soltani2018covert} and \cite{sobers2017covert} used friendly jammers, whereas \cite{he2018covert} examined the secrecy benefits of a Poisson field of interferers.

Recently, Intelligent Reflecting Surfaces (IRS) have emerged as a transformative technology for Sixth-Generation (6G) wireless networks \cite{wu2021intelligent,wu2019towards}, capable of reconfiguring the wireless propagation environment via programmable phase shifts and amplification.
The integration of IRS into covert communication has attracted significant attention.
Early works, such as \cite{deng2022covert}, investigated IRS-assisted relaying, while \cite{lv2021covert,xiao2023star} explored Non-Orthogonal Multiple Access (NOMA) schemes.
The impact of channel uncertainty in IRS-assisted systems was further analyzed in \cite{zou2022irs} and \cite{si2021covert}, showing that the IRS can significantly enhance the covert rate region.
IRS-assisted covert communication with a friendly jammer was studied in \cite{kong2023covert}.

A prominent line of work, exemplified by \cite{si2021covert}, formulates the covert communication problem as a joint optimization of transmit power and IRS phase shifts.
The objective in these studies is typically to maximize the rate subject to a constraint on the detection error probability, i.e., ensuring $\alpha + \beta \ge 1 - \epsilon$ for some small $\epsilon > 0$. While effective, this ``$\epsilon$-covertness'' approach has fundamental limitations.
First, allowing non-zero leakage ($\epsilon > 0$) forces Alice to cap her transmit power at levels comparable to the noise floor.
Second, the resulting optimization problems are often non-convex and computationally intensive.
For instance, the solution in \cite{si2021covert} relies on Semidefinite Relaxation (SDR), which typically entails a high computational complexity scaling as $\mathcal{O}(N^7)$, where $N$ is the number of IRS elements.
This scaling renders real-time implementation infeasible for large surfaces.

Unlike existing approaches, this paper \footnote{Parts of this work were presented to the 2024 58th Asilomar Conference on Signals, Systems, and Computers.} investigates the stricter and theoretically distinct regime of \textbf{perfect covertness}, where we require the signal power at Willie to be exactly zero (i.e., $\alpha + \beta = 1$). This shift in perspective leads to a fundamentally different design paradigm with several key advantages:
\begin{itemize}
    \item \textcolor{black}{Unbounded transmit power (in the ideal perfect-CSI cancellation model) enabled by enforcing zero received energy at the warden.
    By nullifying the signal at Willie under perfect cancellation, Alice is released from the ``low power'' constraint and can, in principle, transmit with arbitrarily high power.
    In practice, residual cancellation uncertainty reintroduces a transmit-power upper bound, later quantified in Theorem~\ref{th:robust_covertness}.}
    \item Unlike \cite{si2021covert}, which requires joint power and phase optimization, our approach optimizes only the IRS phases to achieve signal cancellation.
    \item We propose a gradient-based phase design algorithm with per-iteration complexity $\mathcal{O}(N)$.
    The proposed approach avoids matrix lifting and semidefinite constraints, thereby dramatically reducing computational complexity compared to SDR-based schemes.
\end{itemize}
 
While \cite{zhou2021intelligent,wang2022active} investigated perfect covertness using \emph{active} IRS architectures, which rely on signal amplification to enhance cancellation capability, this work focuses on the more practically scalable regime of \emph{passive} IRS.
Active IRS elements require integrated power amplifiers, thereby increasing hardware complexity, power consumption, and thermal noise. 
In contrast, passive IRS elements are low-cost and energy-efficient.
However, achieving perfect covertness in this regime requires precise multipath interference cancellation without amplification, which fundamentally makes feasibility more challenging.
Thus, we address these practical limitations of achieving `perfect' cancellation by introducing an \emph{operational perfect covertness} model.
Motivated by recent studies on realistic detection \cite{lee2014achieving}, we analyze the system under imperfect CSI and finite detector resolution, ensuring our theoretical gains translate to practical deployments as well.

Beyond the architectural distinction, \cite{zhou2021intelligent} assumes perfect cancellation is feasible under given channel realizations and focuses on delay-constrained transmission design. 
\textcolor{black}{Similarly, \cite{chen2022performance} studies robust IRS-aided covert transmission under hardware impairments, but still considers a KL-based covertness constraint rather than exact warden-signal nulling.}
In contrast, this work characterizes the fundamental feasibility of perfect covertness itself.
In particular, we prove that under standard fading models, the probability of achieving perfect cancellation at the warden converges to one almost surely as the number of passive IRS elements increases.

To contextualize our work within the state of the art, we note that recent literature has continued to expand the IRS covertness landscape. 
For instance, \cite{wang2022covert} explored UAV-mounted IRS architectures to secure air-to-ground links, while \cite{wang2024covert} investigated two-way IRS relaying under noise power uncertainty, showing that bidirectional reflection can further degrade the warden's detection accuracy. 
Similarly, \cite{wu2023irs} analyzed the impact of transmission prior probabilities on covert throughput in IRS-assisted systems.
However, these works largely focus on the $\epsilon$-covertness regime, where a non-zero detection probability is tolerated.
Our work complements this body of knowledge by strictly establishing the fundamental limits and feasibility of the \emph{perfect covertness} (zero-detection) regime.

\textcolor{black}{ Table~\ref{tab:related_work_comparison} summarizes the main differences between this work and representative IRS-assisted covert communication studies. In particular, we distinguish between amplitude-controlled IRS models, in which the reflection magnitude can be adjusted over $[0,1]$, and active IRS models, in which the elements can amplify the incident signal.}

The main contributions of this paper are as follows:
  \textcolor{black}{
    \begin{itemize}
    \item We introduce the concept of \emph{perfect covertness} for IRS-assisted covert communication and establish a necessary and sufficient condition for its achievability. 
    \item For the special case of $N=2$ reflecting elements, we derive a complete closed-form characterization of the entire set of feasible phase configurations.
    \item For general $N$, we distinguish between the full Bob-aware design problem and the feasibility subproblem associated with perfect covertness. 
    For the latter, we formulate the IRS phase design as a warden-signal nulling problem and propose a gradient-based algorithm with per-iteration computational complexity $O(N)$.
    We analyze the optimization landscape induced by the warden-nulling objective and prove that, whenever a perfectly covert configuration exists, gradient descent (GD) with random initialization converges almost surely to a \emph{global} minimizer. 
    Furthermore, under the adopted fading model, we show that a perfectly covert phase configuration exists eventually almost surely as the number of IRS elements increases.
    \item We numerically show that the remaining IRS degrees of freedom can preserve Bob's link after perfect covertness is enforced.
    In particular, as $N$ increases, Bob-aware initialization guides the Willie-nulling GD process toward feasible points whose Bob received power approaches the maximum achievable when the IRS phases are optimized only for Bob, without the Willie-nulling constraint.
    \item We generalize the framework to account for imperfect CSI and finite detector resolution by introducing the notion of \emph{operational perfect covertness}.
    Under a bounded CSI-error model, we derive an explicit sufficient condition on the transmit power that guarantees indistinguishability at the warden in the presence of channel uncertainty. 
    This result yields a deterministic robustness guarantee that bridges the gap between ideal perfect cancellation and practical implementations.
\end{itemize}
}

\begin{table*}[t]
\centering
\caption{\textcolor{black}{Comparison With Representative IRS-Assisted Covert Communication Works}}
\label{tab:related_work_comparison}
\renewcommand{\arraystretch}{1.15}
\setlength{\tabcolsep}{4pt}
\textcolor{black}{
\begin{tabular}{lcccc}
\hline
Related Work & IRS Type & Covertness Criterion & CSI / Robustness & Perfect-Covertness Feasibility \\
\hline
This work & Passive IRS & Perfect Covertness  & Bounded CSI; detector resolution & \checkmark \\
~\cite{zhou2021intelligent} & Amplitude-controlled IRS & KL / perfect achievability & Global CSI; no Willie inst. CSI & Partial \\
~\cite{zou2022irs} & Passive IRS & Detection threshold & Willie channel/noise uncertainty & -- \\ 
~\cite{lv2021covert} & Passive IRS-NOMA & Detection-error probability& Statistical Willie CSI & -- \\
~\cite{wang2022covert} & Passive UAV-IRS & Detection-error probability& Willie location/noise uncertainty & -- \\ 
~\cite{wang2024covert} & Passive two-way IRS & Detection-error probability& Noise uncertainty; random AN & -- \\ 
~\cite{wang2022active} & Active IRS & KL-divergence constraint & Active IRS noise & -- \\ 
~\cite{chen2022performance} & Amplitude-controlled IRS & KL-divergence constraint & Hardware impairments & -- \\
\end{tabular}
}
\end{table*}
\begin{table*}[t]
\centering
\caption{\textcolor{black}{Main Notation}}
\label{tab:notation}
\renewcommand{\arraystretch}{1.15}
\setlength{\tabcolsep}{5pt}
\textcolor{black}{
\begin{tabular}{ll|ll}
\hline
Symbol & Meaning & Symbol & Meaning \\
\hline
$N$ & Number of IRS elements & $\boldsymbol{\phi}$ & IRS phase-shift vector \\ 
$\boldsymbol{v}$ & IRS reflection vector, $v_i=e^{j\phi_i}$ & $\boldsymbol{\Theta}$ & IRS reflection matrix, $\boldsymbol{\Theta}=\operatorname{Diag}(\boldsymbol{v})$ \\
$P_a$ & Alice's transmit power & $\mathcal{D}(\cdot\|\cdot)$ & Kullback--Leibler divergence \\
$h_{a,b}$, $h_{a,w}$ & Alice-to-Bob and Alice-to-Willie channels & $\boldsymbol{h}_{a,s}$ & Alice-to-IRS channel \\
$\boldsymbol{g}_{s,b}$, $\boldsymbol{g}_{s,w}$ & IRS-to-Bob and IRS-to-Willie channels & $\alpha,\beta$ & False-alarm and missed-detection probabilities \\ 
$y_b^1$, $y_w^1$ & Received signals at Bob and Willie under $H_1$ & $P^0_{y_w}$, $P^1_{y_w}$ & Willie output distributions under $H_0$ and $H_1$ \\
$z_i$ & Cascaded Willie-channel coefficient, $z_i=g_{s,w,i}h_{a,s,i}$ & $P_w(\boldsymbol{\phi})$ & Willie received-power objective \\ 
$r(\boldsymbol{\phi})$ & Residual effective channel at Willie & $\eta(\boldsymbol{\phi};N)$ & IRS-reflected channel magnitude toward Willie \\
$T_n$ & Willie energy-detector statistic & $\epsilon_{\mathrm{det}}$ & Detector resolution threshold \\
$e_{a,w}$, $\boldsymbol{e}_{a,s}$, $\boldsymbol{e}_{s,w}$ & Channel-estimation error terms & $\epsilon_w,\epsilon_{a,s},\epsilon_{s,w}$ & Bounds on channel-estimation errors 
\\ \hline
\end{tabular}
}
\end{table*}

\section{System Model and Problem Statement} \label{sec:sysmodel}
Throughout this paper, bold lowercase letters denote vectors (e.g., $\mathbf{h} \in \mathbb{C}^L$) and bold uppercase letters denote matrices (e.g., $\mathbf{H} \in \mathbb{C}^{M \times L}$).
$\mathbb{C}\mathcal{N}(0,\sigma^2 \mathbf{I}_N)$ denotes a circularly symmetric complex Gaussian random vector.
We denote by $\operatorname{Diag}(\mathbf{v})$ a diagonal matrix with the elements of vector $\mathbf{v}$ on the main diagonal.
\textcolor{black}{
For convenience, the main mathematical symbols and abbreviations used throughout the paper are summarized in Table~\ref{tab:notation}.
}
\subsection{System Model} \label{sec:sysmodelSub}
Consider the basic setting of covert communication, in which Alice wants to send a message to Bob without being detected by Willie.
Alice uses a passive IRS with $N$ elements.
Alice has $M$ transmitting antennas, while Bob and Willie have $K$ and $L$ receiving antennas, respectively.

\begin{figure}[!t]
\centering
\includegraphics[width=0.8\linewidth]{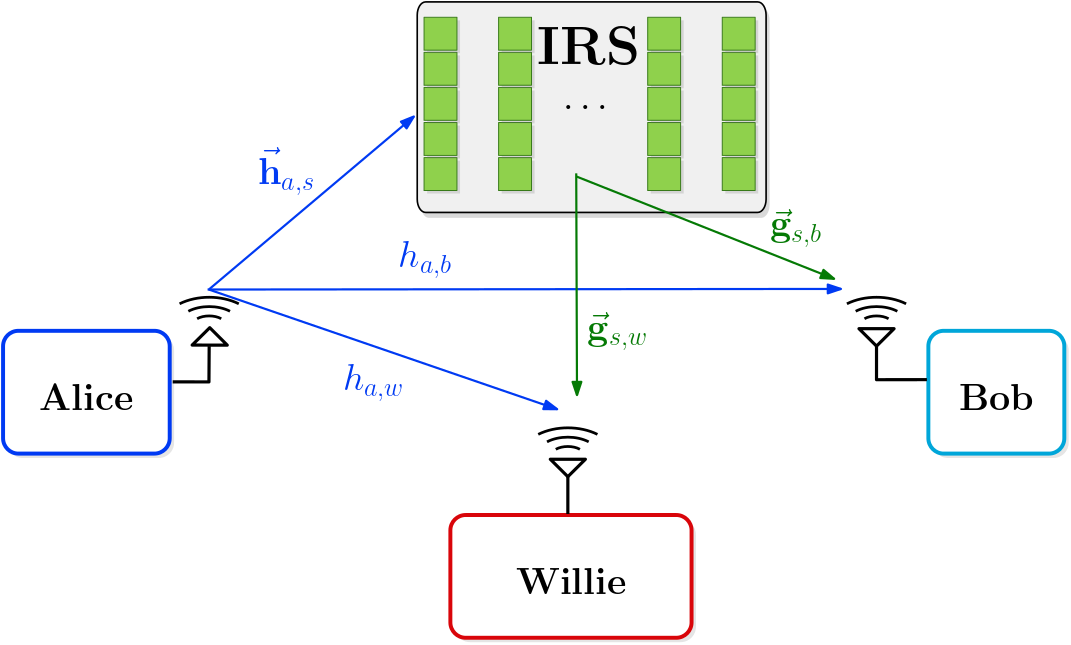}
\caption {IRS-Assisted Covert Communication.}
\label{fig:generalcase}
\end{figure} 

In this work, we simplify the problem and focus on the case where all participants have a single antenna (i.e., $K=1$, $L=1$, $M=1$), yet with $N$ elements in the IRS (see Fig.~\ref{fig:generalcase}).
This enables us to focus on IRS optimization, the paper's key novelty.
We examine the received signals in Bob and Willie's channels under two hypotheses.
The null hypothesis, $H_0$, assumes no transmission, and $H_1$ assumes Alice transmitted.
Let $y^1_b$ and $y^1_w$ denote the received signals at Bob and Willie under $H_1$, respectively, assuming $x$ was transmitted.
Under a single-antenna model at all parties and an N-array IRS, we have
\begin{equation}\label{eq:system_model}
    \begin{split}
        &y_b^1 = \left( \mathbf{g}_{s,b}^T \boldsymbol{\Theta} \, \mathbf{h}_{a,s}+h_{a,b}\right)x+n_b\\
        &y_w^1 = \left( \mathbf{g}_{s,w}^T \boldsymbol{\Theta} \, \mathbf{h}_{a,s}+h_{a,w}\right)x+n_w
    \end{split}
\end{equation}
where $\boldsymbol{\phi} = [\phi_1, \ldots, \phi_N]^T \in [0,2\pi)^N$ 
denotes the IRS phase-shift vector.
The corresponding unit-modulus reflection coefficient vector is defined as $\boldsymbol{v} = [e^{j\phi_1}, \ldots, e^{j\phi_N}]^T$, and the IRS reflection matrix is given by $\boldsymbol{\Theta} = \operatorname{Diag}(\boldsymbol{v}) \in \mathbb{C}^{N \times N}$.
Here, $\phi_n$ represents the controllable 
phase response of the $n$th IRS element.
$n_b\thicksim\mathbb{C}\mathcal{N}(0,{\sigma}^2_{b})$ and
$n_w\thicksim\mathbb{C}\mathcal{N}(0,{\sigma}^2_{w})$
are independent, representing Bob's and Willie's channel noises, respectively.
$\mathbf{h}_{a,s}\thicksim\mathbb{C}\mathcal{N}(\mathbf{0},\sigma_{a,s}^2 \mathbf{I}_N)$ is a channel coefficient vector between Alice and the IRS with i.i.d entries.
$h_{a,b}\thicksim\mathbb{C}\mathcal{N} (0,\sigma_{a,b}^2)$ and
$h_{a,w}\thicksim\mathbb{C}\mathcal{N} (0,\sigma_{a,w}^2)$
are the channel coefficients from Alice to Bob and from Alice to Willie, respectively.
${\mathbf{g}_{s,w}\thicksim\mathbb{C}\mathcal{N}(\mathbf{0},\sigma_{s,w}^2  \mathbf{I}_N)}$, and
${\mathbf{g}_{s,b}\thicksim\mathbb{C}\mathcal{N}(\mathbf{0},\sigma_{s,b}^2  \mathbf{I}_N)}$ are channel coefficient vectors from the IRS to Willie and Bob, respectively, each with i.i.d entries. 
All channel coefficients are mutually independent.
The received signal-to-noise ratios (SNR) at Bob and Willie under $H_1$ are:
\begin{equation*}
SNR_b^1=\frac{P_a{\abs{\mathbf{g}_{s,b}^T \boldsymbol{\Theta} \, \mathbf{h}_{a,s}+h_{a,b}}}^2}{\sigma_b^2}
\end{equation*}
\begin{equation*}\label{eq:SNRw}
SNR_w^1=\frac{P_a {\abs{\mathbf{g}_{s,w}^T \boldsymbol{\Theta} \, \mathbf{h}_{a,s}+h_{a,w}}^2}}{\sigma_w^2},
\end{equation*}
where $P_a$ is Alice's transmit power.

\subsection{Perfect Covertness}
Let $P_{y_w}^0$ and $P_{y_w}^1$ denote Willie's output distributions.
Under $H_0$, the only energy observed by Willie's receiver is the noise, and therefore $P_{y_w}^0 = n_w\thicksim\mathbb{C}\mathcal{N}\left(0,\sigma_w^2\right)$.
However, under $H_1$, Alice's signal propagates through both the direct channel and the IRS.

\begin{definition}[Perfect Covertness]
A perfect covertness solution is achieved when 
$\mathcal{D}(P_{y_w}^0||P_{y_w}^1) =\mathcal{D}(\mathbb{P}(y_w|H_0)||\mathbb{P}(y_w|H_1)) = 0$,
where $\mathcal{D}(\cdot||\cdot)$ denotes the Kullback-Leibler divergence.
\end{definition}
In other words, Willie's channel has identical signal distributions under $H_0$ (the null hypothesis) and $H_1$ (the alternative hypothesis).
The definition is motivated by the fact that \cite{lehmann2005testing}
\begin{equation*}\label{eq:vardistAlphaBeta}
    \alpha + \beta = 1-\mathcal{V}(P_{y_w}^0,P_{y_w}^1),
\end{equation*}
where $\alpha$ is the probability of mistakenly accepting $H_1$ and $\beta$ is the probability of mistakenly accepting $H_0$.
$\mathcal{V}(P_{y_w}^0,P_{y_w}^1)$ is the total variation distance between $P_{y_w}^0$ and
$P_{y_w}^1$.
Note that by Pinsker’s inequality \cite[Lemma 11.6.1]{cover1999elements}:
\begin{equation*}\label{eq:Pinsker}
    \mathcal{V}(P_{y_w}^0,P_{y_w}^1)\leq\sqrt{\frac{1}{2}\mathcal{D}(P_{y_w}^0||P_{y_w}^1)}.
\end{equation*}
Thus,  $\alpha +\beta=1$ iff $\mathcal{D}(P_{y_w}^0||P_{y_w}^1) = 0$.
The following lemma asserts that perfect covertness is achieved if and only if Willie’s SNR is zero.
\begin{lemma}\label{lemma:SNRtoPerfect}
   $SNR_w^1 = 0$ iff  $\mathcal{D}(P_{y_w}^0||P_{y_w}^1) = 0$.
\end{lemma}

\begin{IEEEproof}
Suppose $SNR_w^1 = 0$, which means that either $P_a=0$ or ${\abs{\mathbf{g}_{s,w}^T \boldsymbol{\Theta} \, \mathbf{h}_{a,s}+h_{a,w}}=0}$, hence the received signal at Willie's antenna is pure noise and $P_{y_w}^0 = P_{y_w}^1$.
For the other direction, suppose $\mathcal{D}(P_{y_w}^0||P_{y_w}^1) = 0$. 
Then, $P_{y_w}^0=P_{y_w}^1$ almost everywhere.
Since $y_w^1$ is zero-mean complex Gaussian with variance $\sigma_w^2+P_a{\abs{\mathbf{g}_{s,w}^T \boldsymbol{\Theta} \, \mathbf{h}_{a,s}+h_{a,w}}}^2$, equality with $P_{y_w}^0$ implies $P_a{\abs{\mathbf{g}_{s,w}^T \boldsymbol{\Theta} \, \mathbf{h}_{a,s}+h_{a,w}}}^2=0$ and hence $SNR_w^1=0$.
\end{IEEEproof}

We can now formally define our problem.
Most prior work selects $(P_a,\boldsymbol{\phi})$ to satisfy an epsilon-covertness constraint $\mathcal{D}(P_{y_w}^0||P_{y_w}^1)<\epsilon$.
Here, we impose the stronger requirement
${\mathcal{D}(P_{y_w}^0||P_{y_w}^1) = 0}$.

\begin{equation}\label{eq:optimization_problem}
    \begin{aligned}
        \max_{{\boldsymbol{\phi}} \in [0,2\pi)^N}\quad & {\abs{\mathbf{g}_{s,b}^T \boldsymbol{\Theta} \, \mathbf{h}_{a,s}+h_{a,b}}}^2\\
        \textrm{s.t.} \quad & {\abs{\mathbf{g}_{s,w}^T \boldsymbol{\Theta} \, \mathbf{h}_{a,s}+h_{a,w}}^2}=0.
    \end{aligned}
\end{equation}

The goal is to find a phase vector $\boldsymbol{\phi}$ that nulls Willie’s SNR completely, resulting in  $\alpha+\beta = 1$, while maximizing Bob's SNR.
In the ideal perfect-cancellation model, Alice's transmission power is unconstrained by the covertness constraint. 
Thus, Alice can transmit at maximum power. 
While Lemma \ref{lemma:SNRtoPerfect} provides the condition for \textit{perfect covertness}, the following lemma tells under which \emph{channel coefficients} this condition is possible to achieve. 
\begin{lemma}\label{lemma:generalCond}
The perfect covertness condition,
$\left|\mathbf{g}_{s,w}^T \boldsymbol{\Theta} \, \mathbf{h}_{a,s}+h_{a,w}\right|=0$
is satisfied iff 
\begin{equation}\label{eq:condition}
    \underset{\boldsymbol{\phi}}{\min} \ \eta(\boldsymbol{\phi};N)\leq\left|h_{a,w}\right|\leq \underset{\boldsymbol{\phi}}{\max} \ \eta(\boldsymbol{\phi};N)
\end{equation}
\textcolor{black}{where $\eta(\boldsymbol{\phi};N) \overset{\Delta}{=} \ \left|{\mathbf{g}_{s,w}^T \boldsymbol{\Theta} \, \mathbf{h}_{a,s}}\right|$ is the reflected-path magnitude.}
\end{lemma}
\begin{IEEEproof}
Suppose $\left|h_{a,w}\right|$ satisfies the condition in Lemma~\ref{lemma:generalCond}.
We can simplify $\eta(\boldsymbol{\phi};N)$ as follows:
\begin{equation*}
    \begin{split}
        \eta(\boldsymbol{\phi};N)&=\left|\mathbf{g}_{s,w}^T \boldsymbol{\Theta} \, \mathbf{h}_{a,s}\right|
        =\left|\sum_{i=1}^{N} z_i e^{j\phi_i}\right|,
    \end{split}
\end{equation*}
where $z_i=g_{{s,w}_i} h_{{a,s}_i}$, $i\in\{1,\ldots,N\}$.
$\eta(\boldsymbol{\phi};N)$ is a continuous function with a connected domain.
According to the intermediate value theorem, for any $\tau\in\left[ \underset{\boldsymbol{\phi}}{\min} \ \eta(\boldsymbol{\phi};N), \  \underset{\boldsymbol{\phi}}{\max} \ \eta(\boldsymbol{\phi};N) \right]$ 
there exists $\boldsymbol{\phi}_0$ such that $\eta(\boldsymbol{\phi}_0;N) = \tau$.
In particular, by our assumption there exists $\boldsymbol{\phi}_0$ such that 
$\eta(\boldsymbol{\phi}_0;N)=|h_{a,w}|.$
Additionally, note that $\eta$ is invariant under a global phase rotation,
$$\eta(\phi_1,\ldots,\phi_N;N)=\eta(\phi_1+\phi,\ldots,\phi_N+\phi;N) \quad , \phi\in [0,2\pi].$$
Thus, a specific $\phi\in[0,2\pi]$ can be chosen to rotate $\mathbf{g}_{s,w}^T \boldsymbol{\Theta} \, \mathbf{h}_{a,s}$ in the opposite direction of $h_{a,w}$, leading to
$\left|\mathbf{g}_{s,w}^T \boldsymbol{\Theta} \, \mathbf{h}_{a,s}+h_{a,w}\right|=0 $.

For the reverse direction, let us first denote $\angle z$ as the angle between the real axis and the complex number $z$.  
Assume there exist a phase vector $\boldsymbol{\phi}_0$ such that
$$\left|\mathbf{g}_{s,w}^T \boldsymbol{\Theta} \, \mathbf{h}_{a,s}+h_{a,w}\right|=0. $$

\noindent Then,
$\angle(\mathbf{g}_{s,w}^T \boldsymbol{\Theta} \, \mathbf{h}_{a,s})= \angle h_{a,w} - \pi \textrm{ and } \eta(\boldsymbol{\phi}_0;N) = \left|{\mathbf{g}_{s,w}^T \boldsymbol{\Theta} \, \mathbf{h}_{a,s}}\right| =\left|h_{a,w}\right|$
which clearly confirms that the condition in Lemma~\ref{lemma:generalCond} is satisfied.
\end{IEEEproof}

\textcolor{black}{\begin{remark}
The Rayleigh fading model is adopted in this paper to obtain explicit probabilistic feasibility results. 
However, the perfect covertness condition itself is deterministic and applies to any fixed channel realization.
In particular, the condition
\begin{equation*}
 {\abs{\mathbf{g}_{s,w}^T \boldsymbol{\Theta} \, \mathbf{h}_{a,s}+h_{a,w}}^2}=0
\end{equation*}
does not rely on the Rayleigh distribution. 
Therefore, the same nulling condition holds for alternative statistical channel models, such as Rician fading. 
In this case, the distributions of the channel coefficients differ from those in the previously considered model. 
\end{remark} }

\section{Perfect Covertness with $N=2$ IRS Elements}\label{sec:N=2}

We first study the perfect-covertness feasibility and phase design problem under perfect CSI for the case of $N=2$ IRS elements.

\subsection{\textcolor{black}{Instantaneous Willie CSI}}
We assume that Alice has full channel state information (CSI) for the link to Willie. This setting models an internal adversary: Willie is a legitimate network participant, yet Alice seeks to keep the transmission covert from Willie.

From \eqref{eq:optimization_problem}, the objective is to maximize Bob's SNR while nullifying Willie's SNR.
\begin{figure}[!t]
     \centering
    \includegraphics[width=0.9\linewidth]{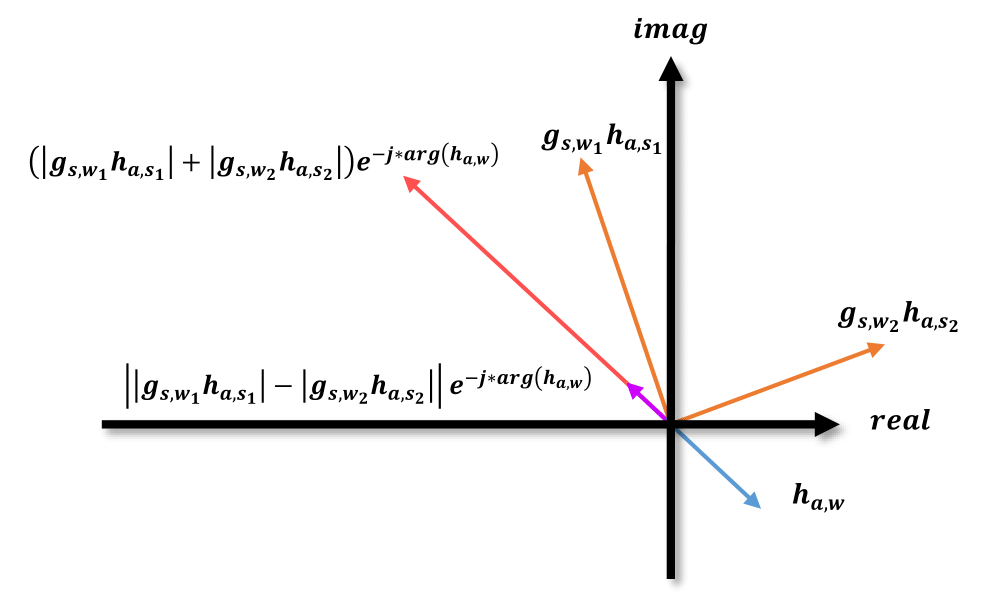}
     \caption{Geometrical visualization of the Perfect Covertness solution for the case of $N=2$ IRS elements.}
     \label{fig:geoIntu}
\end{figure} 

As outlined in the system model, the IRS controller controls the phase shifts.
Considering the simple case of $N=2$ IRS elements (see Fig.~\ref{fig:geoIntu}), it is intuitive to understand how the phase shifts can be adjusted to nullify Willie's SNR.
In Fig. \ref{fig:geoIntu}, the orange vectors represent the two complex components of the indirect path from Alice and Willie.
The black vector represents the channel coefficient of the direct channel from Alice to Willie.
The phase shifts $\phi_1,\phi_2$ must be determined so that the resulting vector after rotation (i.e., ${{g_{s,w}}_1 {h_{a,s}}_1   e^{j \phi_1}+ {g_{s,w}}_2  {h_{a,s}}_2  e^{j \phi_2}}$) matches the magnitude but is opposite in direction to the direct channel vector, thus creating destructive interference. 
This results in a zero SNR at Willie’s receiver.

To this end, we first obtain the maximum and minimum achievable values of the effective channel magnitude:
\textcolor{black}{
\begin{equation*}\label{eq:maxminfor2}
    \begin{split}
        &\underset{\phi_1,\phi_2}{\max} \ \eta(\boldsymbol{\phi};2) =\left|{g_{s,w}}_1 {h_{a,s}}_1\right|+\left|{g_{s,w}}_2 {h_{a,s}}_2\right|,\\
        &\underset{\phi_1,\phi_2}{\min} \ \eta(\boldsymbol{\phi};2) =\big|\left|{g_{s,w}}_1 {h_{a,s}}_1\right| - \left|{g_{s,w}}_2 {h_{a,s}}_2\right|\big|.
    \end{split}
\end{equation*}
}

The following lemma provides a complete characterization of all feasible solutions to (\ref{eq:optimization_problem}) in the $N=2$ case.
Define $z_i \triangleq {g_{s,w}}_i {h_{a,s}}_i$ and 
$\phi_{z_i} \triangleq \angle z_i$.
\begin{lemma} \label{lemma:solutinN2}
Suppose the feasibility condition from Lemma~\ref{lemma:generalCond} is satisfied.
Let $\Delta \triangleq \angle z_2 - \angle z_1$ and define $\gamma \triangleq \cos^{-1}\!\left( \frac{|h_{a,w}|^2 - |z_1|^2 - |z_2|^2}{2|z_1||z_2|} \right)$.
The set of phases $\phi_1 \in [0, 2\pi)$ satisfying $|z_1 e^{j\phi_1} + z_2| = |h_{a,w}|$ is given by
\begin{equation}
\phi_1 = \Delta \pm \gamma \pmod{2\pi}.
\end{equation}
\end{lemma}

\begin{IEEEproof}
The condition $|z_1 e^{j\phi_1} + z_2| = |h_{a,w}|$ implies that the vectors $z_1 e^{j\phi_1}$ and $z_2$ sum to a vector of magnitude $|h_{a,w}|$, forming a triangle with side lengths $|z_1|$, $|z_2|$, and $|h_{a,w}|$. \textcolor{black}{The relative angle between $z_1e^{j\phi_1}$ and $z_2$ is $ \psi = (\angle z_1+\phi_1)-\angle z_2=\phi_1-\Delta .$}
By the Law of Cosines, $|h_{a,w}|^2 = |z_1|^2 + |z_2|^2 + 2|z_1||z_2|\cos\psi$. Isolating $\cos\psi$ yields $\cos(\phi_1 - \Delta) = \cos\gamma$, from which the result follows directly.
\end{IEEEproof}

The solution procedure for \eqref{eq:optimization_problem} proceeds as follows.
First, the two candidate values $\phi_1^{(1)}$ and $\phi_1^{(2)}$ satisfying
\[
\left|\mathbf{g}_{s,w}^T \boldsymbol{\Theta} \, \mathbf{h}_{a,s}\right| = |h_{a,w}|
\]
are obtained by applying Lemma~\ref{lemma:solutinN2}.
Next, for each candidate $\phi_1^{(i)}$, a common phase offset
\[
\delta^{(i)}
= -\angle\!\left(z_1 e^{j\phi_1^{(i)}} + z_2\right)
+ \angle h_{a,w} + \pi
\]
is applied to both $z_1 e^{j\phi_1^{(i)}}$ and $z_2$, ensuring that
\[
\angle\!\left(\mathbf{g}_{s,w}^T \boldsymbol{\Theta} \, \mathbf{h}_{a,s}\right)
= \angle h_{a,w} + \pi .
\]
Finally, among the resulting configurations, the one that maximizes Bob’s SNR is selected.


We emphasize that the solution described above is valid only when the condition in Lemma~\ref{lemma:generalCond} is satisfied.
This condition, in turn, depends on the random channel coefficients.
\textcolor{black}{
Simulation results reveal that the probability of attaining perfect covertness when \(N = 2\) is rather limited. In the next section, we analytically show that by increasing the number of IRS elements, this probability can be made arbitrarily close to 1, thus guaranteeing the existence of a perfectly covert solution with high probability.
}

Define the following random variables,
\begin{equation}\label{eq:definitionsDoubleR}
\begin{split}
&X_1 = \left|h_{{a,s}_1}\right|\left|g_{{s,w}_1}\right|
\sim \text{Double Rayleigh}(\sigma_{x_1}),\\
&X_2 = \left|h_{{a,s}_2}\right|\left|g_{{s,w}_2}\right|
\sim \text{Double Rayleigh}(\sigma_{x_2}),\\
&Y = \left|h_{a,w}\right|
\sim \text{Rayleigh}(\sigma_y),
\end{split}
\end{equation}
where $\sigma_{x_1} = \sigma_{a,s}\sigma_{s,w}$,
$\sigma_{x_2} = \sigma_{a,s}\sigma_{s,w}$, and
$\sigma_y = \sigma_{a,w}$ denote the corresponding scale parameters.

The following lemma characterizes the probability that this condition holds.
\begin{lemma}[Feasibility Probability] \label{lemma:probability_calc}
Let $X_1, X_2, Y$ denote the random magnitudes defined in \eqref{eq:definitionsDoubleR}. The probability that a perfectly covert solution exists is given by
\begin{align} \label{eq:prob_integral}
P_{\text{feas}} &= 2 \int_{0}^{\infty} f_Y(y) \Bigg[ \int_{0}^{y/2} f_{X_2}(x_2) \int_{y-x_2}^{y+x_2} f_{X_1}(x_1) \, dx_1 \, dx_2 \nonumber \\
&+ \int_{y/2}^{\infty} f_{X_2}(x_2) \int_{x_2}^{y+x_2} f_{X_1}(x_1) \, dx_1 \, dx_2 \Bigg] dy.
\end{align}
\end{lemma}

\begin{IEEEproof}
From Lemma~\ref{lemma:solutinN2}, a solution exists if $|X_1 - X_2| \le Y \le X_1 + X_2$. Since $X_1$ and $X_2$ are i.i.d., the probability is symmetric with respect to $X_1$ and $X_2$. Thus, we compute the probability for the case $X_1 \ge X_2$ and multiply by 2:
\begin{equation}
P_{\text{feas}} = 2 \mathbb{P}(X_1 - X_2 \le Y \le X_1 + X_2, X_1 \ge X_2).
\end{equation}
Conditioning on $Y=y$ and $X_2=x_2$, the inequality for $X_1$ becomes $\max(x_2, y-x_2) \le X_1 \le y+x_2$. The lower bound $\max(x_2, y-x_2)$ behaves differently depending on whether $x_2 < y/2$ or $x_2 \ge y/2$. Splitting the integration over $x_2$ at $y/2$ yields the expression in \eqref{eq:prob_integral}.
\end{IEEEproof}

\begin{figure}[!t]
     \centering
     \begin{tikzpicture}
        \node[inner sep=0] (image) at (0,0) {\includegraphics[width=\linewidth]{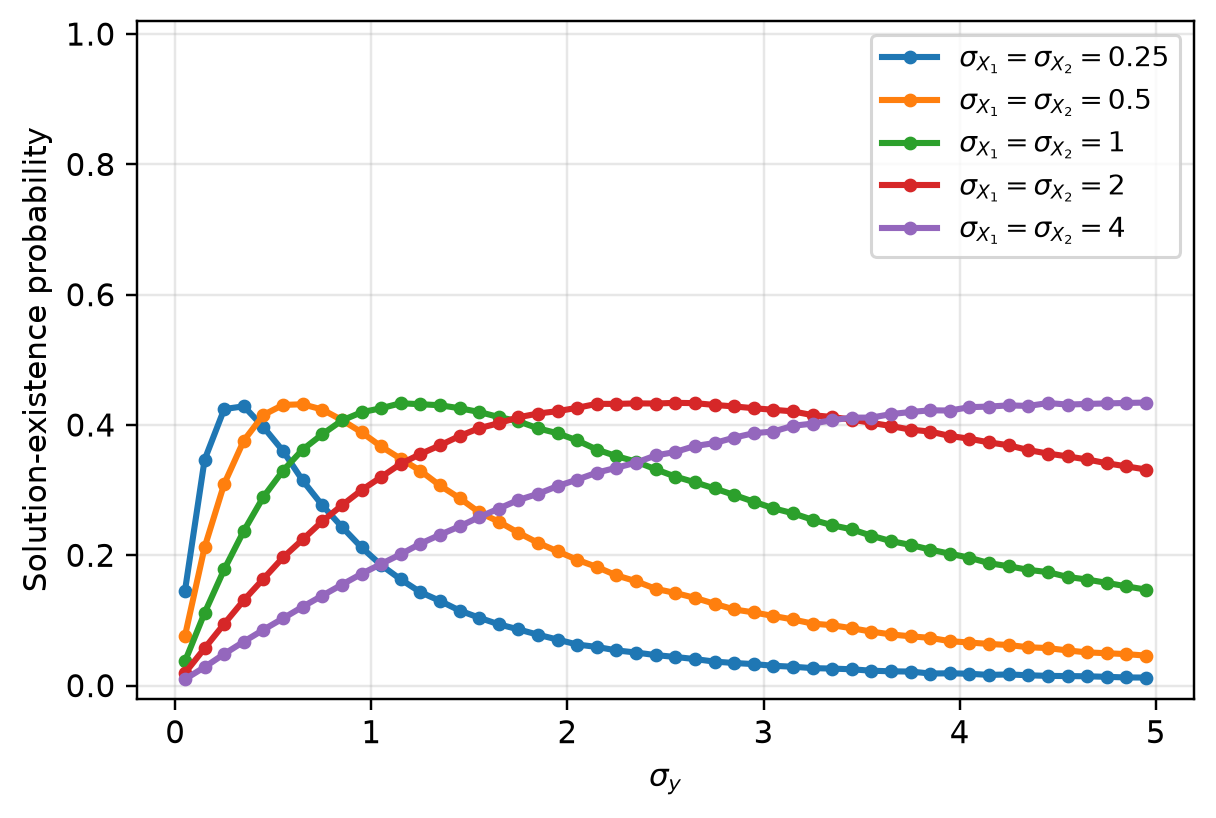}};
    \end{tikzpicture}
     \caption{Probability of solution existence versus $\sigma_y$ for different $\sigma_{X_1}=\sigma_{X_2}$, where $N=2$ IRS elements.}
     \label{fig:numericProbEqualSigma}
\end{figure}

Fig.~\ref{fig:numericProbEqualSigma} illustrates the probability of solution existence as a function of $\sigma_y$ for several values of $\sigma_{X_1}=\sigma_{X_2}$, assuming $N=2$ IRS elements.
The maximum probability is approximately $0.43$.
Moreover, as $\sigma_{X_1}=\sigma_{X_2}$ increases, the range of $\sigma_y$ for which a solution exists broadens, leading to an overall increase in the probability.

These results indicate that achieving perfect covertness with only two IRS elements is inherently challenging, as the probability of satisfying the feasibility condition does not approach $1$.
Nevertheless, the $N=2$ case provides valuable insight into the structure of the problem.
As shown in Section~\ref{sec:N>2}, increasing the number of IRS elements allows the probability of achieving perfect covertness to become arbitrarily close to $1$, thereby enabling perfect covertness with high probability.

\section{Perfect Covertness with  $N>2$ IRS Elements}\label{sec:N>2}
We now extend the analysis to the case of an IRS with more than two reflecting elements.
Unlike the $N=2$ case studied in the previous section, an explicit closed-form characterization of the IRS phase configuration that achieves perfect cancellation at Willie is no longer available when $N>2$.

\textcolor{black}{
Nevertheless, we prove that the perfect-covertness condition is satisfied eventually almost surely as the number of IRS elements grows. 
Equivalently, under the considered fading model, the probability that a perfectly covert IRS phase configuration exists converges to one as $N\to\infty$.
We further propose an efficient GD-based algorithm that converges to such a solution whenever a feasible configuration exists.
}

Finally, simulation results confirm the analytical findings and show that perfect covertness can be achieved with high probability even for moderate IRS sizes, with $N$ as small as $8$.

\subsection{Probability Analysis}\label{sec:probAnal}
We analyze the probability that a phase configuration satisfying the perfect covertness condition exists, i.e., that the condition in Lemma~\ref{lemma:generalCond} is satisfied.
Define the event
\[
S_N \triangleq
\left\{
\min_{\boldsymbol{\phi}} \eta(\boldsymbol{\phi};N)
\le |h_{a,w}|
\le \max_{\boldsymbol{\phi}} \eta(\boldsymbol{\phi};N)
\right\}.
\]

\begin{theorem}\label{theorem:solutionExistanceN->inf}
Under the channel model of Section~II,
\[
\mathbb{P}\big(\exists N_0:\ \forall N\ge N_0,\ S_N\big)=1.
\]
That is, the perfect covertness condition is satisfied eventually almost surely
as the number of IRS elements $N$ grows.
\end{theorem}
For a detailed proof, refer to Appendix~\ref{app:solutionExistanceN->inf}.

Motivated by Theorem~\ref{theorem:solutionExistanceN->inf}, we next develop an efficient method for finding an IRS phase vector that satisfies the perfect-covertness condition.

\subsection{A Perfectly Covert IRS Phase-Design Scheme}

\textcolor{black}{Problem ~\eqref{eq:optimization_problem} contains two components: a strict feasibility constraint at Willie and an objective corresponding to Bob's received power. 
Since perfect covertness requires exact satisfaction of the nulling constraint, we first focus on the feasibility component of \eqref{eq:optimization_problem}, namely, finding an IRS phase-shift vector that drives Willie's effective channel to zero.
The feasible set may contain phase configurations with different Bob received powers.
In the numerical section, we therefore use Bob-aware initialization to guide the Willie-nulling GD process toward feasible points with stronger Bob received power.
Accordingly, Algorithm~\ref{Schem-algo} is designed to solve the warden-nulling subproblem.}

\textcolor{black}{We propose an iterative phase-design procedure based on GD to solve  
\begin{equation}\label{eq:relaxedProb}
\min_{\boldsymbol{\phi}\in[0,2\pi)^N}
P_w(\boldsymbol{\phi})
=
\left|\sum_{i=1}^N z_i e^{j\phi_i}+h_{a,w}\right|^2 ,
\end{equation}
and we establish that this procedure converges almost surely to a global minimizer. In particular, if there exist phase vectors that drive Willie’s received power to zero, i.e., if there exists \(\boldsymbol{\phi}\) such that \(P_w(\boldsymbol{\phi})=0\), then the proposed scheme converges almost surely to one such phase vector.
}

\textcolor{black}{Algorithm~\ref{Schem-algo} takes as input the coefficients
$\{z_i\}_{i=1}^N$ and the scalar $h_{a,w}$, and applies GD
directly to $P_w(\boldsymbol{\phi})$ over all IRS phases
$\boldsymbol{\phi}\in[0,2\pi)^N$.}

\begin{algorithm}[t]
{
\caption{Gradient-Based IRS Phase Design}
\label{Schem-algo}
\textcolor{black}{
\begin{algorithmic}[1]
\Require $\{z_i\}_{i=1}^{N}$, $h_{a,w}$, step size $s$, tolerance $\delta$, maximum iterations $I_{\max}$
\Ensure IRS phase vector $\boldsymbol{\phi}$
\State Initialize $\boldsymbol{\phi}^{(0)}\in[0,2\pi)^N$
\For{$i=0,1,\ldots,I_{\max}-1$}
    \State $r^{(i)} \gets \sum_{k=1}^{N} z_k e^{j\phi_k^{(i)}}+h_{a,w}$
    \For{$k=1,\ldots,N$}
        \State $\left[\nabla P_w(\boldsymbol{\phi}^{(i)})\right]_k
        \gets
        -2\operatorname{Im}
        \left\{
        z_k e^{j\phi_k^{(i)}}\left(r^{(i)}\right)^*
        \right\}$
    \EndFor
    \State $\boldsymbol{\phi}^{(i+1)}
    \gets
    \left(
    \boldsymbol{\phi}^{(i)}
    -
    s\nabla P_w(\boldsymbol{\phi}^{(i)})
    \right)
    \operatorname{mod} 2\pi$
    \If{$\left|P_w(\boldsymbol{\phi}^{(i+1)})-P_w(\boldsymbol{\phi}^{(i)})\right|\leq\delta$}
        \State \Return $\boldsymbol{\phi}^{(i+1)}$
    \EndIf
\EndFor
\State \Return $\boldsymbol{\phi}^{(I_{\max})}$
\end{algorithmic}
}
}
\end{algorithm}

\textcolor{black}{\textit{Computational Complexity and Initialization:}
As shown in Appendix~\ref{appendix:gradient}, each iteration of Algorithm~\ref{Schem-algo} first computes the residual
\[
r^{(i)}
=
\sum_{k=1}^{N} z_k e^{j\phi_k^{(i)}}+h_{a,w},
\]
which requires $\mathcal{O}(N)$ operations. 
Once $r^{(i)}$ is available, all entries of $\nabla P_w(\boldsymbol{\phi}^{(i)})$ are evaluated in closed form by a single additional pass over the IRS elements, which also requires $\mathcal{O}(N)$ operations.
Therefore, the per-iteration computational complexity of Algorithm~\ref{Schem-algo} is $\mathcal{O}(N)$.
The worst-case complexity is $\mathcal{O}(NI_{\max})$, and for a fixed number of iterations, the complexity grows linearly with the number of IRS elements.}

\textcolor{black}{In practice, as shown in Sec.~\ref{sec:numerical_results}, initializing the phase vector by beamforming toward Bob significantly accelerates convergence and consistently yields better performance than random initialization.}
\textcolor{black}{
\begin{theorem}\label{theorem:GloOptConvergence}
Consider fixed coefficients $\{z_i\}_{i=1}^{N}$, with $z_i\neq 0$ for all $i$, and the objective
\[
P_w(\boldsymbol{\phi})
=
\left|
\sum_{i=1}^{N} z_i e^{j\phi_i}+h_{a,w}
\right|^2.
\]
Assume that gradient descent is initialized from a uniform distribution over $[0,2\pi)^N$ and uses a sufficiently small fixed step size. 
Then, the iterates converge to a global minimizer of $P_w(\boldsymbol{\phi})$ with probability one.
\end{theorem}
We note that if a perfectly covert phase vector exists, then the global minimum of $P_w(\boldsymbol{\phi})$ is zero.
In this case, Theorem~\ref{theorem:GloOptConvergence} implies that the algorithm converges to a phase vector satisfying $P_w(\boldsymbol{\phi})=0$. 
The convergence result is deterministic once the channel coefficients are fixed and therefore does not depend on the particular fading distribution used to generate them.}
\subsubsection{Proof of Convergence}
The GD algorithm converges to a critical point $\boldsymbol{\phi}^*$ satisfying $\nabla f(\boldsymbol{\phi}^*)=\mathbf{0}$.
This critical point can be a local minimum or a saddle point.

We first consider the auxiliary objective obtained by setting $h_{a,w}=0$ and define $S(\boldsymbol{\phi})\triangleq\sum_{i=1}^{N} z_i e^{j\phi_i}$.
Later, we generalize it to the case $h_{a,w}\neq0$.

The proof relies on recent results on strict saddle functions.
In particular,
\cite[Theorem~4.1]{lee2016gradient} establishes that gradient descent with random initialization almost surely avoids saddle points when the objective function is a strict saddle.
We therefore characterize the critical points of $|S(\boldsymbol{\phi})|^2$ and establish its strict saddle structure.

\begin{definition}[Strict Saddle]\label{definition:strictSaddle}
A critical point $\boldsymbol{\phi}^*$ of $f$ is a strict saddle
if $\lambda_{min} (\nabla^2 f(\boldsymbol{\phi}^*)) < 0$.
\end{definition}
Intuitively, strict saddle points require at least one direction along which the curvature is strictly negative.
\begin{definition}[Strict Saddle Function]
A function $f$ is a strict saddle function if the Hessian matrix of every saddle point has a negative eigenvalue.
\end{definition}

The following lemma characterizes the critical points of $|S(\boldsymbol{\phi})|^2$.
\begin{lemma}[Critical Point Characterization] \label{lemma:180degree}
Let $c_k(\boldsymbol{\phi}) \triangleq \sum_{i\neq k} z_i e^{j\phi_i}$ denote the \textbf{partial sum of the remaining $N-1$ elements}. A phase vector $\boldsymbol{\phi}^*\in[0,2\pi)^N$ is a critical point of $|S(\boldsymbol{\phi})|^2$ if and only if for all $k\in\{1,\ldots,N\}$, the $k$-th term is collinear with this partial sum, i.e.,
\begin{equation}
\angle\left(z_k e^{j\phi^*_k}\right) = \angle\left(c_k(\boldsymbol{\phi}^*)\right) + n\pi, \quad n\in \mathbb{Z}.
\end{equation}
\end{lemma}

\begin{IEEEproof}
We isolate the dependence on the $k$-th phase $\phi_k$ by writing
\begin{equation}
\begin{split}
|S(\boldsymbol{\phi})|^2 &= |z_k e^{j\phi_k} + c_k|^2  
\\&
=|z_k|^2 + |c_k|^2 + 2|z_k||c_k|\cos(\phi_k + \angle z_k - \angle c_k).    
\end{split}
\end{equation}
Here, $\phi_k + \angle z_k = \angle(z_k e^{j\phi_k})$.
The partial derivative is $[\nabla f]_k = -2|z_k||c_k|\sin(\angle(z_k e^{j\phi_k}) - \angle c_k)$. A point is critical iff $[\nabla f]_k = 0$ for all $k$, which implies $\sin(\cdot) = 0$. Thus, the phase difference between the $k$-th vector and the partial sum $c_k$ must be an integer multiple of $\pi$.
\end{IEEEproof}

The next lemma argues that any critical point that does not attain the global minimum
corresponds to full alignment of all vectors along a single axis.
\begin{lemma}[Alignment of Non-Optimal Critical Points] \label{lemma:critical aligned}
Let $\boldsymbol{\phi}^*$ be a critical point of $|S(\boldsymbol{\phi})|^2$ that is not a global minimum (i.e., $|S(\boldsymbol{\phi}^*)|^2 > 0$). Then, all constituent vectors $z_i e^{j\phi^*_i}$ share a common axis, i.e.,
\begin{equation}
\angle (z_i e^{j\phi^*_i}) = \angle (z_l e^{j\phi^*_l}) + k\pi, \quad \forall i,l \in \{1,\ldots,N\}, k\in \mathbb{Z}.
\end{equation}
\end{lemma}

\begin{IEEEproof}
Let $v_k \triangleq z_k e^{j\phi^*_k}$ be the $k$-th signal component. 
Thus, $S(\boldsymbol{\phi}) \triangleq \sum_{i=1}^N v_i$.
By Lemma~\ref{lemma:180degree}, at any critical point $\boldsymbol{\phi}^*$, $v_k$ is collinear with the partial sum $c_k = S(\boldsymbol{\phi}^*) - v_k$.
Mathematically, this implies $v_k = \alpha_k (S(\boldsymbol{\phi}^*) - v_k)$ for some real scalar $\alpha_k$. 
Rearranging terms yields $v_k(1+\alpha_k) = \alpha_k S(\boldsymbol{\phi}^*)$. 
Since the solution is not a global minimum, we have $|S(\boldsymbol{\phi}^*)|^2  > 0$.
Additionally, as the channel gains are almost surely non-zero, $v_k \neq 0$. 
It follows that every vector $v_k$ is a scalar multiple of the same non-zero vector $S(\boldsymbol{\phi})$. Therefore, all vectors are collinear.
\end{IEEEproof}

Finally, the curvature of $|S(\boldsymbol{\phi})|^2$ at such points implies the strict saddle property.
\begin{lemma}[Strict Saddle Property] \label{lemma:saddlemax}
All critical points of $|S(\boldsymbol{\phi})|^2$ that do not correspond to the global minimum are strict saddle points.
\end{lemma}
The proof is given in  Appendix \ref{Appendix:Lemma8}.

\begin{corollary} \label{cor:PWRisStrict}
The objective function $|S(\boldsymbol{\phi})|^2$ is a strict saddle function.
\end{corollary}
\begin{IEEEproof}
A function is a strict saddle function if every critical point is either a local minimum or a strict saddle point (i.e., has a strictly negative Hessian eigenvalue).
Lemma~\ref{lemma:saddlemax} establishes precisely this property for all non-optimal critical points of $|S(\boldsymbol{\phi})|^2$.
\end{IEEEproof}
Using Lemmas \ref{lemma:180degree}, \ref{lemma:critical aligned}, and \ref{lemma:saddlemax}, we can now prove Theorem \ref{theorem:GloOptConvergence}.

\begin{IEEEproof}[Proof of Theorem \ref{theorem:GloOptConvergence}]
Corollary~\ref{cor:PWRisStrict} confirms that $|S(\boldsymbol{\phi})|^2$ satisfies the strict saddle property. Furthermore, as shown in Appendix~\ref{appendix:Lipschitz}, the gradient $\nabla |S(\boldsymbol{\phi})|^2$ is Lipschitz continuous. Consequently, the objective satisfies the conditions of \cite[Theorem~4.1]{lee2016gradient}, which guarantees that gradient descent with random initialization converges to a local minimizer almost surely. 
Since Lemma~\ref{lemma:saddlemax} implies there are no spurious local minima (all local minima are global), the algorithm converges to a perfect covert solution almost surely.
\end{IEEEproof}

\paragraph*{Extension to the Full Objective}
The results established for the auxiliary objective $|S(\boldsymbol{\phi})|^2$ extend to the full objective $\textcolor{black}{P_w(\boldsymbol{\phi})} = |S(\boldsymbol{\phi}) + h_{a,w}|^2$ through the same collinearity structure. 
Specifically, the stationarity condition, $\textcolor{black}{\nabla P_w} = \mathbf{0}$, dictates that every reflected component $z_k e^{j\phi_k}$ \textcolor{black}{is collinear with $c_k(\boldsymbol{\phi})+h_{a,w}$.
This is geometrically identical to the condition derived in Lemma \ref{lemma:180degree}, with the partial sum $c_k(\boldsymbol{\phi})$  replaced by $c_k(\boldsymbol{\phi})+h_{a,w}$.}
Consequently, any non-optimal critical point of the full objective retains the collinear structure of the auxiliary case.
As in Lemma \ref{lemma:saddlemax}, this alignment is structurally unstable: a rotational perturbation exists that breaks the alignment and reduces the error, thereby preserving strict saddle properties. 

\subsection{\textcolor{black}{Discussion on Multi-Warden and Multi-User Extensions}}
\textcolor{black}
{
The core analysis in this work considers a single legitimate receiver and a warden equipped with one antenna, though several extensions are feasible. 
In particular, the proposed framework can be directly generalized to settings involving multiple wardens.
In such scenarios, ensuring perfect covertness requires that the effective received signal be simultaneously canceled at every warden.
This model can equally describe a single warden equipped with multiple receiving antennas.
In fact, if the wardens fully cooperate, then they are mathematically equivalent to a single multi-antenna warden.
}

\textcolor{black}{
Compared with the single-antenna case, this imposes $L$ complex cancellation constraints on the same IRS phase vector.
Therefore, feasibility becomes more restrictive and depends on the ratio of IRS elements to warden antennas.
}

\textcolor{black}{
The extension to multiple legitimate receivers is more dependent on the intended service model.
If the receivers cooperate, the Bob-side objective can be chosen as a joint received power.
If different receivers decode distinct messages, the design must account for user-specific rates and tradeoffs among them.
We therefore keep the analytical development focused on one legitimate receiver and leave a complete multi-user formulation for future work.
}

\textcolor{black}{
Yet, the numerical results in Fig.~\ref{fig:multi_willie} show that the proposed phase-design method can succeed in practice and considerably reduce the received signal power at a multi-antenna warden.
This suggests that the approach remains useful beyond the single-antenna model, although a complete theoretical characterization of the multi-antenna case is left for future work.
}

\section{Imperfect CSI and Detection Resolution} 
\label{sec:resolution}

\textcolor{black}{The preceding analysis assumes perfect CSI and ideal IRS implementation, allowing exact cancellation of the effective channel toward the warden and thus $\mathrm{SNR}_w^1=0$.
In practice, however, the direct link $h_{a,w}$ and the cascaded IRS channel $(\mathbf{h}_{a,s},\mathbf{g}_{s,w})$ are known only up to bounded estimation errors.
Moreover, practical IRS hardware impairments, such as phase errors, finite-resolution phase shifts, and amplitude mismatches, may also create a residual mismatch between the designed and the actual effective Alice-IRS-Willie channel. 
The key question, therefore, is whether perfect covertness can withstand such realistic residual uncertainties. 
In this section, we show that it remains operationally attainable, provided that the resulting residual signal lies below the warden's minimum resolvable energy level.}

We assume that Willie employs an energy detector, a natural and widely adopted model in the covert communication literature 
\cite{bash2014lpd}. 
When Gaussian signaling is used, and the codebook is unknown to the warden, the binary hypothesis test reduces to detecting a variance shift, for which the optimal Neyman-Pearson test depends only on the received energy \cite{bash2014lpd}. 
Accordingly, over a block of $n$ channel uses, Willie observes $\{y_w[k]\}_{k=1}^n$ as in \eqref{eq:system_model} and computes
\begin{equation*}
T_n \triangleq \frac{1}{n}\sum_{k=1}^n |y_w[k]|^2.
\end{equation*}

Let $r$ denote the implemented residual channel, ${r\triangleq\mathbf{g}_{s,w}^T \boldsymbol{\Theta} \mathbf{h}_{a,s} + h_{a,w}}$.
Then, under hypotheses $H_0$ and $H_1$, we have
\begin{equation*}
\mathbb{E}[T_n \mid H_0] = \sigma_w^2, \quad
\mathbb{E}[T_n \mid H_1] = \sigma_w^2 + P_a |r|^2,
\end{equation*}
so that the mean shift induced by Alice’s transmission is given by
\begin{equation*}
\Delta_T \triangleq
\mathbb{E}[T_n \mid H_1] - \mathbb{E}[T_n \mid H_0]
= P_a |r|^2.
\label{eq:delta_T}
\end{equation*}

\subsection{Detector Resolution and Operational Covertness}

Practical radiometers exhibit finite sensitivity due to quantization, calibration errors, and hardware limitations.
We model these impairments via a deterministic resolution threshold $\epsilon_{\mathrm{det}}>0$.

\begin{definition}[Detector Resolution]
 A detector has energy resolution $\epsilon_{\mathrm{det}}>0$ if changes in the mean energy statistic smaller than $\epsilon_{\mathrm{det}}$ cannot be reliably resolved.
\end{definition}

\begin{definition}[Operational Perfect Covertness]
A transmission is operationally perfectly covert if
\[
\Delta_T \le \epsilon_{\mathrm{det}},
\]
so that the detector cannot significantly outperform random guessing.
\end{definition}

\begin{figure}[!t]
    \centering
    \includegraphics[width=\linewidth]{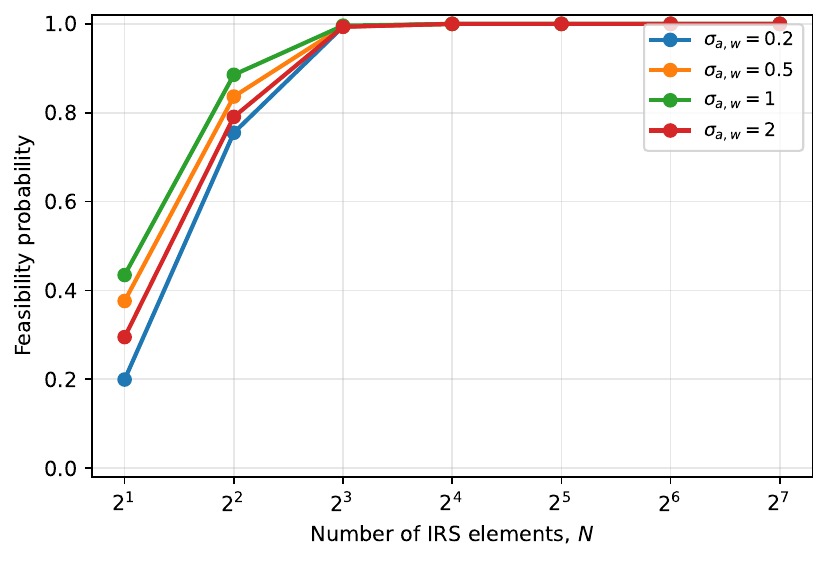}
    \caption{\textcolor{black}{Probability that perfect covertness is feasible versus the number of IRS elements.}}
    \label{fig:feasibility_vs_n}
\end{figure}

\subsection{Bounded CSI Uncertainty Model}

To ensure perfect operational covertness, the system must robustly suppress the signal below $\epsilon_{\mathrm{det}}$ despite channel estimation errors \textcolor{black}{and implementation-induced residual mismatches}.

We adopt a bounded error model where the IRS controller has access to estimates $\hat{h}_{a,w}$, $\hat{\mathbf{h}}_{a,s}$, and $\hat{\mathbf{g}}_{s,w}$. 
We write
\begin{align}
h_{a,w}      &= \hat{h}_{a,w} + e_{a,w}, \label{eq:csi_err1}\\
\mathbf{h}_{a,s} &= \hat{\mathbf{h}}_{a,s} + \mathbf{e}_{a,s}, \label{eq:csi_err2}\\
\mathbf{g}_{s,w} &= \hat{\mathbf{g}}_{s,w} + \mathbf{e}_{s,w}, \label{eq:csi_err3}
\end{align}

where the error terms are bounded by known constants:
\begin{equation}
    |e_{a,w}| \le \epsilon_w, \quad
    \|\mathbf{e}_{a,s}\|_\infty \le \epsilon_{a,s}, \quad
    \|\mathbf{e}_{s,w}\|_\infty \le \epsilon_{s,w}.
    \label{eq:error_bounds}
\end{equation}

The following theorem provides a sufficient condition for covertness that unifies CSI uncertainty bounds with the transmit power constraint.

\begin{theorem}[Robust Operational Covertness Condition]\label{th:robust_covertness}
    Assume the bounded CSI error model in \eqref{eq:error_bounds}. 
    Let the IRS phase matrix $\boldsymbol{\Theta}$ be designed to perfectly cancel the estimated effective channel, i.e., $\hat{\mathbf{g}}_{s,w}^T \boldsymbol{\Theta} \hat{\mathbf{h}}_{a,s} + \hat{h}_{a,w} = 0$. 
    The transmission is guaranteed to be operationally perfectly covert provided the transmit power $P_a$ satisfies:
    \begin{equation}
        P_a \le \frac{\epsilon_{\mathrm{det}}}{\delta_{\mathrm{CSI}}^2},
        \label{eq:power_bound}
    \end{equation}
    where $\delta_{\mathrm{CSI}}$ represents the worst-case residual channel magnitude:
    \begin{equation*}
        \delta_{\mathrm{CSI}} \triangleq \sqrt{N}\epsilon_{a,s}\|\hat{\mathbf{g}}_{s,w}\|_2 + \sqrt{N}\epsilon_{s,w}\|\hat{\mathbf{h}}_{a,s}\|_2 + N\epsilon_{s,w}\epsilon_{a,s} + \epsilon_w.
        \label{eq:delta_csi}
    \end{equation*}
\end{theorem}

\begin{figure*}[!t]
    \centering
    \includegraphics[width=0.9\textwidth]{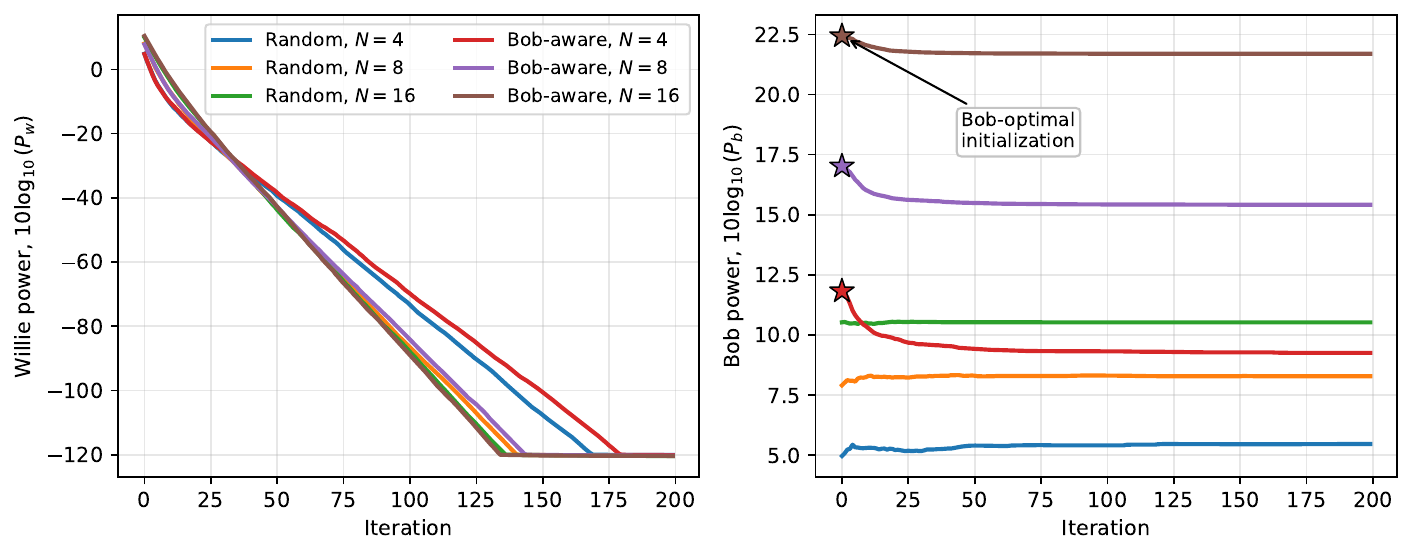}
       \caption{\textcolor{black}{Median GD convergence for Willie nulling and Bob received power.
       Stars mark the Bob-aware iteration-$0$ points, which are Bob-unrestricted coherent-combining optima before Willie-nulling GD begins.}}
    \label{fig:gd_convergence}
\end{figure*}

\begin{IEEEproof}
    Substituting the channel models into the effective channel expression yields:
    \begin{equation*}
        \begin{split}
        r &= (\hat{\mathbf{g}}_{s,w} + \mathbf{e}_{s,w})^T \boldsymbol{\Theta} (\hat{\mathbf{h}}_{a,s} + \mathbf{e}_{a,s}) + (\hat{h}_{a,w} + e_{a,w})  \\
          &= \underbrace{(\hat{\mathbf{g}}_{s,w}^T \boldsymbol{\Theta} \hat{\mathbf{h}}_{a,s} + \hat{h}_{a,w})}_{=0}  + \hat{\mathbf{g}}_{s,w}^T \boldsymbol{\Theta} \mathbf{e}_{a,s}  
          \\&
          \qquad  \qquad +
          \mathbf{e}_{s,w}^T \boldsymbol{\Theta} \hat{\mathbf{h}}_{a,s} + \mathbf{e}_{s,w}^T \boldsymbol{\Theta} \mathbf{e}_{a,s} + e_{a,w}. 
        \end{split}
    \end{equation*}
    Assuming the estimated channels admit a perfectly covert solution per Lemma \ref{lemma:generalCond}, the phase matrix $\boldsymbol{\Theta}$ is configured such that the first term vanishes.
    We bound the magnitude of the remaining terms using the triangle inequality and the Cauchy-Schwarz inequality ($\|\mathbf{x}^T \mathbf{y}\| \le \|\mathbf{x}\|_2 \|\mathbf{y}\|_2$). Noting that the spectral norm of the IRS phase matrix is $\|\boldsymbol{\Theta}\|_2 = 1$:
    \begin{equation*}
        \begin{split}
        |r| &\le \|\hat{\mathbf{g}}_{s,w}\|_2 \|\mathbf{e}_{a,s}\|_2 + \|\mathbf{e}_{s,w}\|_2 \|\hat{\mathbf{h}}_{a,s}\|_2
        \\&
        \qquad\qquad  + 
        \|\mathbf{e}_{s,w}\|_2 \|\mathbf{e}_{a,s}\|_2 + |e_{a,w}|. 
        \end{split}
    \end{equation*}
    We convert the $\ell_\infty$ error bounds from \eqref{eq:error_bounds} to $\ell_2$ bounds using the inequality $\|\mathbf{x}\|_2 \le \sqrt{N}\|\mathbf{x}\|_\infty$ for $\mathbf{x} \in \mathbb{C}^N$. Thus, $\|\mathbf{e}_{a,s}\|_2 \le \sqrt{N}\epsilon_{a,s}$ and $\|\mathbf{e}_{s,w}\|_2 \le \sqrt{N}\epsilon_{s,w}$. Substituting these into the inequality:
    \begin{equation*}
        |r| \le \delta_{\mathrm{CSI}}.
    \end{equation*}
    The induced mean power shift is therefore bounded by ${\Delta_T = P_a |r|^2 \le P_a \delta_{\mathrm{CSI}}^2}$.
    Imposing the condition $P_a \le \epsilon_{\mathrm{det}} / \delta_{\mathrm{CSI}}^2$ ensures $\Delta_T \le \epsilon_{\mathrm{det}}$, satisfying the definition of operational perfect covertness.
\end{IEEEproof}

\textcolor{black}{
\subsection{Random CSI Uncertainty Model}
The bounded-uncertainty framework can also be used to obtain a probabilistic guarantee in the presence of random channel estimation errors. Define the CSI uncertainty event
\[
\mathcal{E}_{\mathrm{CSI}}
=
\left\{
|e_{a,w}|\leq \epsilon_w,\,
\|\mathbf{e}_{a,s}\|_\infty\leq \epsilon_{a,s},\,
\|\mathbf{e}_{s,w}\|_\infty\leq \epsilon_{s,w}
\right\},
\]
where the error terms \(e_{a,w}\), \(\mathbf{e}_{a,s}\), and
\(\mathbf{e}_{s,w}\) are random with known probability distributions.
Let \(\rho\in[0,1]\) denote a prescribed outage probability.
 If the uncertainty bounds \(\epsilon_w\), \(\epsilon_{a,s}\), and \(\epsilon_{s,w}\) are chosen such that
\[
\Pr(\mathcal{E}_{\mathrm{CSI}})\geq 1-\rho,
\]
then Theorem~\ref{th:robust_covertness} applies on the event \(\mathcal{E}_{\mathrm{CSI}}\).
Therefore, under the same transmit-power constraint $
P_a \leq \epsilon_{\mathrm{det}}/{\delta_{\mathrm{CSI}}^2},
$
we obtain
\[
\Pr\left(\Delta_T\leq \epsilon_{\mathrm{det}}\right)\geq 1-\rho.
\]
Thus, the transmission is operationally perfectly covert with probability at least \(1-\rho\). 
In this sense, the deterministic worst-case guarantee provided by
Theorem~\ref{th:robust_covertness} is converted into a high-probability
operational covertness guarantee.
}

\section{\textcolor{black}{Numerical Results}}
\label{sec:numerical_results}
\textcolor{black}{
In this section, we provide numerical results to support the analytical findings and evaluate the proposed IRS phase-design method. 
We begin by detailing the simulation setup.
The subsequent simulations are structured to examine: (i) the feasibility condition for achieving perfect covertness, (ii) the convergence behavior of the proposed gradient descent (GD) algorithm, (iii) Bob's retained received power relative to the unconstrained Bob-power reference, (iv) robustness when Willie is equipped with multiple antennas, and (v) imperfect CSI.}

\subsection{\textcolor{black}{Simulation Setup}}
\textcolor{black}{
 All results were generated from fixed random seeds. 
The code and configuration files used to generate the numerical results are publicly available in the accompanying repository~\cite{elimelech2026perfectlycovertcode}. 
 All channel coefficients are independent circularly symmetric complex Gaussian random variables with unit average power, i.e.,  $\sigma_{a,s}=\sigma_{s,w}=\sigma_{s,b}=\sigma_{a,w}=\sigma_{a,b}=1$. 
 The IRS phase shifts are continuous and have unit modulus.
 The step size is chosen according to the Lipschitz constant of $\nabla P_w(\boldsymbol{\phi})$.
 Specifically, let $L_w$ be any constant satisfying
\begin{equation*}
    \|\nabla^2 P_w(\boldsymbol{\phi})\|_2 \leq L_w,
    \qquad \forall \boldsymbol{\phi}\in[0,2\pi)^N .
\end{equation*}
This bound implies that $\nabla P_w(\boldsymbol{\phi})$ is $L_w$-Lipschitz, which is the standard smoothness condition ensuring sufficient descent with fixed-step gradient descent. 
Thus, choosing a step size smaller than the inverse of the Lipschitz constant keeps each GD update in the stable descent regime~\cite{nesterov2004introductory}. 
As shown in Appendix~\ref{appendix:Lipschitz}, one possible choice is
\begin{equation}
    L_w = 4\max_i \left(|z_i|\sum_{m\neq i}|z_m|\right)
    +2|h_{a,w}|\max_i |z_i|.
\end{equation}
In the simulations, we use $s=\alpha/L_w$, with $\alpha=1$.
The stopping tolerance is $10^{-12}$ and the success threshold is $10^{-10}$ unless otherwise noted.
}

{\color{black}
\subsection{Feasibility of Perfect Covertness}
We first verify the exact geometric feasibility condition.
Fig.~\ref{fig:feasibility_vs_n} shows the resulting feasibility probability over $10^4$ Monte Carlo trials for $N\in\{2,4,8,16,32,64,128\}$ and several direct Alice-Willie channel strengths.
The feasibility probability grows quickly with $N$, supporting Theorem \ref{theorem:solutionExistanceN->inf}, which states that the perfect covertness condition holds with probability one as the number of IRS elements increases.

\subsection{Convergence of Willie Nulling}

Fig.~\ref{fig:gd_convergence} reports the convergence behavior of the proposed GD method for $N\in\{4,8,16\}$ using $1000$ feasible channel realizations and a maximum of $200$ iterations.
Two initializations are considered: random uniform phases and Bob-aware initialization, where the initial phase vector is chosen to coherently combine Bob's channel before applying Willie-nulling GD.
The figure shows the median Willie and Bob powers over the Monte Carlo trials. Stars in the Bob-power panel mark the Bob-aware iteration-$0$ points, which correspond to the Bob-unrestricted coherent-combining optimum before Willie-nulling updates begin. 
The Willie detection power decreases to the numerical floor for both initialization strategies, whereas the Bob-aware initialization remains almost completely unaffected.

\subsection{Bob Performance Under Perfect Covertness}

To quantify the legitimate-link cost of enforcing perfect covertness, we compare the final Bob power with the Bob-unrestricted coherent-combining baseline.
The retained Bob power ratio is
\begin{equation*}
    \rho_b =
    \frac{P_b(\boldsymbol{\phi}_{\mathrm{final}})}
    {P_b(\boldsymbol{\phi}_{b}^{\mathrm{unres}})}.
\end{equation*}
Fig.~\ref{fig:init_benefit} shows the median and $10$th--$90$th percentile range of $10\log_{10}\rho_b$.
Random initialization suffers a growing Bob-power loss as $N$ increases, while Bob-aware initialization keeps the final Bob power close to the Bob-unrestricted benchmark.
For example, at $N=128$, the median loss is approximately $-20.36$ dB with random initialization, but only $-0.071$ dB with Bob-aware initialization.
This confirms that Bob-aware initialization directly improves the operating point obtained after perfect-covertness optimization.

Fig.~\ref{fig:init_benefit} also shows a clear difference in dispersion.
With random initialization, the $10$th--$90$th percentile range remains wide as $N$ increases, indicating that different Willie-nulling solutions can yield very different Bob gains.
With Bob-aware initialization, this range narrows substantially, and the final solution concentrates near the Bob-unrestricted benchmark.

\begin{remark}
Fig.~\ref{fig:init_benefit} illustrates the role of the remaining IRS degrees of freedom after enforcing perfect covertness.
The Willie-nulling condition imposes a fixed cancellation constraint, whereas the number of IRS phases grows with $N$.
Thus, for large $N$, many perfectly covert configurations remain available for improving Bob's link.
However, choosing among them is important.
Random initialization may converge to a feasible point with poor Bob gain, while Bob-aware initialization guides GD toward feasible points that nearly preserve the Bob-unrestricted coherent-combining gain.
\end{remark}

\begin{figure}[!t]
    \centering
    \includegraphics[width=0.9\linewidth]{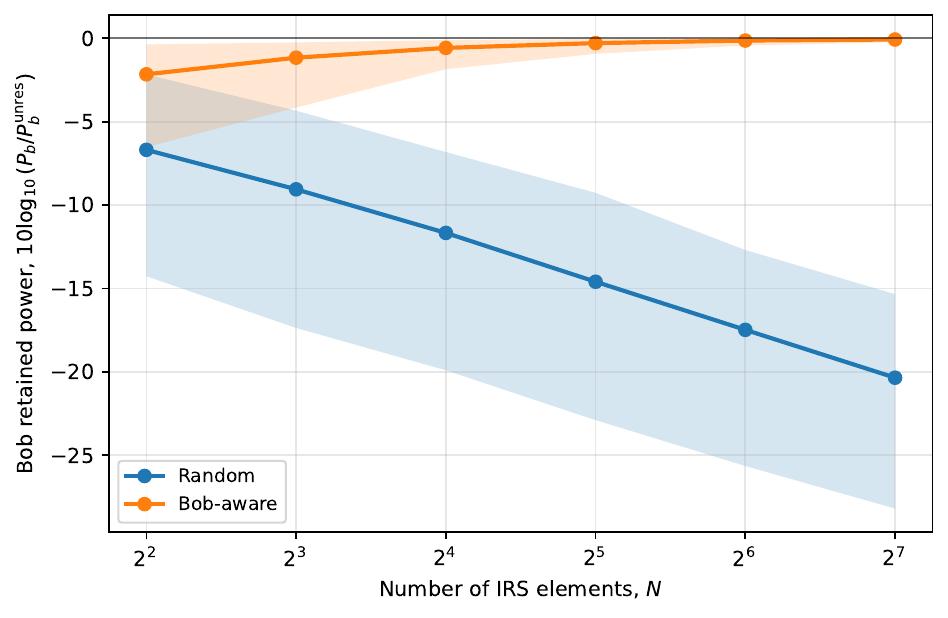}
    \caption{\textcolor{black}{Effect of initialization on the final Bob operating point.
    The plot shows the retained Bob-power ratio $10\log_{10}\rho_b$ after Willie-nulling GD, where $\rho_b=P_b(\boldsymbol{\phi}_{\mathrm{final}})/P_b(\boldsymbol{\phi}_{b}^{\mathrm{unres}})$.}}
    \label{fig:init_benefit}
\end{figure}
\subsection{Multi-Antenna Willie}

We next evaluate an extension in which Willie has $L$ antennas.
For this simulation, let $\mathbf{h}_{a,w}\in\mathbb{C}^{L}$ denote the direct Alice--Willie channel vector and let $\mathbf{G}_{s,w}\in\mathbb{C}^{L\times N}$ denote the IRS--Willie channel matrix.
The Alice--IRS channel remains $\mathbf{h}_{a,s}\in\mathbb{C}^{N}$.
The residual vector at Willie is then

\begin{equation}
    \mathbf{r}_w(\boldsymbol{\phi})
    =
    \mathbf{G}_{s,w}
    \left(\mathbf{h}_{a,s}\odot e^{j\boldsymbol{\phi}}\right)
    + \mathbf{h}_{a,w},
\end{equation}
where $\odot$ denotes element-wise multiplication. 
The objective becomes $P_w^{(L)}(\boldsymbol{\phi})=\|\mathbf{r}_w(\boldsymbol{\phi})\|_2^2$.
Fig.~\ref{fig:multi_willie} shows that increasing $N$ relative to $L$ reduces the final Willie power and improves the success probability.
This supports the practical conclusion that larger IRS arrays can suppress multi-antenna Willie leakage.

\begin{figure*}[!t]
    \centering
    \includegraphics[width=0.9\textwidth]{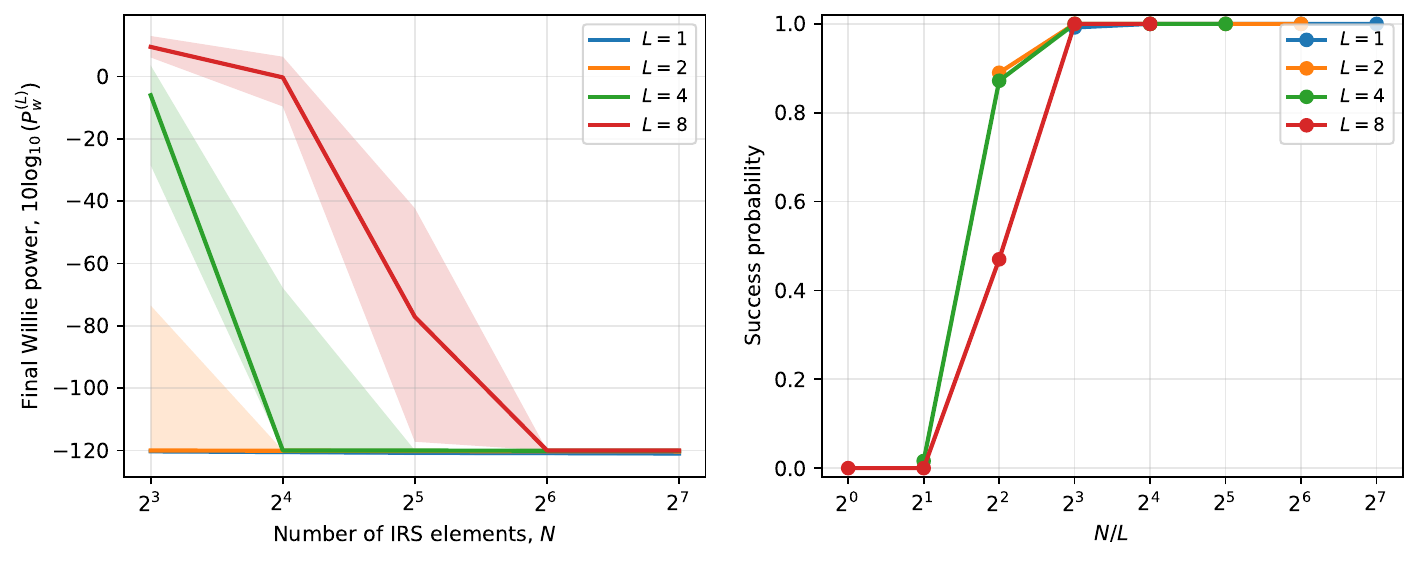}
    \caption{\textcolor{black}{Multi-antenna ($L \ge 1$) Willie experiment.
    The left panel shows the final Willie power after GD, and the right panel shows the success probability.
    Increasing the IRS dimension relative to the number of Willie antennas improves leakage suppression and success probability.}}
    \label{fig:multi_willie}
\end{figure*}

\subsection{Imperfect CSI and Operational Perfect Covertness}

To evaluate robustness to channel-estimation errors, the IRS phases are designed from estimated channels and then tested on perturbed true channels.
Each channel coefficient is perturbed by an independent circularly symmetric complex Gaussian error with variance $\sigma_e^2$.
For each perturbed channel realization, we evaluate the operational-covertness residual $r$ and compute $\Delta_T=P_a |r|^2$.
Operational perfect covertness is declared when $\Delta_T\leq \epsilon_{\mathrm{det}}$.
Figure~\ref{fig:imperfect_csi} uses $N=8$, $5000$ Monte Carlo trials, $\epsilon_{\mathrm{det}}=5\times10^{-4}$, $P_a\in\{0.1,1,10\}$, and Gaussian CSI-error variances $\sigma_e^2\in[10^{-4},10^{-1}]$.
The probability of satisfying the detector-resolution constraint decreases as either the CSI-error variance or transmit power increases.

\begin{figure}[!t]
    \centering
    \includegraphics[width=\linewidth]{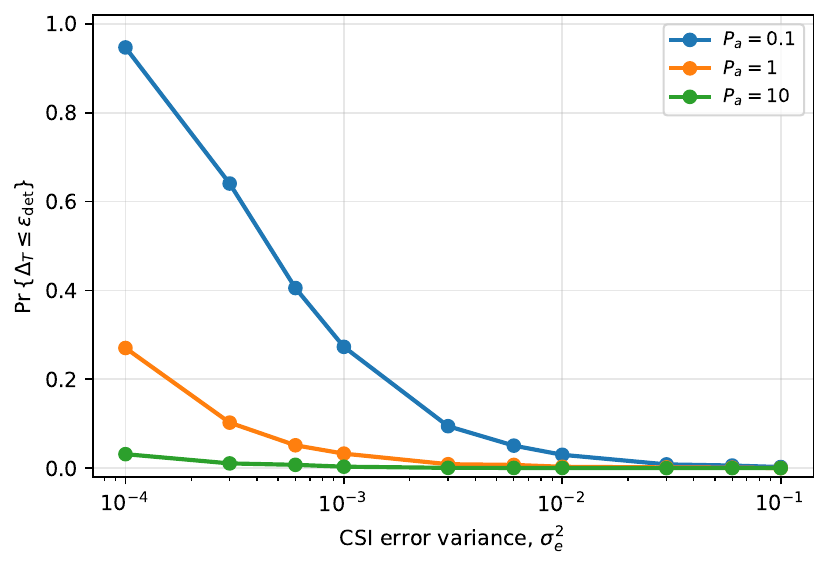}
    \caption{\textcolor{black}{Operational perfect covertness under imperfect CSI.
    Larger estimation errors and higher transmit powers reduce the probability that the residual leakage remains below the detector-resolution threshold.}}
    \label{fig:imperfect_csi}
\end{figure}

Overall, the numerical results support four main conclusions.
First, perfect covertness becomes increasingly feasible as the IRS dimension grows.
Second, the proposed GD method reliably drives Willie leakage to the numerical floor in feasible single-antenna settings, and Bob-aware initialization is essential to preserve the legitimate link.
Third, the proposed framework remains relevant when extended to practical scenarios, such as multi-antenna systems.
Fourth, operational perfect covertness is sensitive to residual channel mismatch, so practical implementations require sufficiently accurate CSI or a transmit-power limit that accounts for the residual uncertainty.
}

\section{Conclusion}
In this work, we investigated the fundamental limits of perfect covert communication assisted by an IRS.
We characterized the necessary and sufficient conditions for feasibility, provided a complete analytical solution for the two-element case, and established that for general array sizes, the perfect-covertness condition is eventually satisfied almost surely as the number of elements increases.

To enable practical implementation, we formulated the phase design as an interference minimization problem.
A key contribution of this work was the rigorous analysis of the optimization landscape.
We proved that the objective function satisfies the \textit{strict saddle property}, ensuring that it is free of spurious local minima. 
Consequently, we showed that standard gradient descent with random initialization converges almost surely to a global minimum and achieves perfect cancellation whenever feasible.
\textcolor{black}{The numerical results further showed that, as the IRS dimension increases, Bob-aware initialization preserves Bob's link and approaches the Bob-unrestricted coherent-combining benchmark while driving Willie leakage to the numerical floor.}
\textcolor{black}{They also showed that larger IRS dimensions improve leakage suppression in the multi-antenna Willie setting.}

Finally, we addressed the physical limitations of channel uncertainty. 
By adopting a bounded CSI error model, we derived robust conditions for \textit{operational perfect covertness}, guaranteeing that the warden's detection capability remains effectively nullified even when exact signal cancellation is mathematically impossible.
\textcolor{black}{The imperfect-CSI results also confirmed that operational perfect covertness is sensitive to residual channel mismatch, especially as the transmit power or CSI-error variance increases.}

\textcolor{black}{
From a practical perspective, the proposed framework is relevant to low-probability-of-detection links in tactical, emergency, and privacy-sensitive wireless systems, where concealing the transmission may be as important as protecting the transmitted content. 
Passive IRS deployment is especially attractive in such scenarios due to its low power consumption and simple hardware structure. 
Several directions remain open beyond the scope of this work, including a complete feasibility characterization for multi-antenna or cooperative wardens, and multi-user Bob-side objectives.
}


\appendices

\section{Proof Of Theorem~\ref{theorem:solutionExistanceN->inf}}
\label{app:solutionExistanceN->inf}
\begin{IEEEproof}
We decompose $S_N=L_N\cap U_N$, where
$${L_N \triangleq
\left\{\min_{\boldsymbol{\phi}} \eta(\boldsymbol{\phi};N)\le |h_{a,w}|\right\}},$$ $${U_N \triangleq
\left\{|h_{a,w}|\le \max_{\boldsymbol{\phi}} \eta(\boldsymbol{\phi};N)\right\}}.
$$

\emph{Upper bound.}
For any fixed realization of $\{z_i\}$, the maximum of
$\eta(\boldsymbol{\phi};N)$ is achieved by phase alignment, yielding
\[
\max_{\boldsymbol{\phi}} \eta(\boldsymbol{\phi};N)
=
\sum_{i=1}^{N} |z_i|.
\]
The random variables $\{|z_i|\}$ are i.i.d., nonnegative, with
$\mathbb{E}[|z_i|]=\mu_z>0$ and $\mathrm{Var}(|z_i|)<\infty$.
By the Strong Law of Large Numbers,
\[
\sum_{i=1}^{N} |z_i| \xrightarrow[]{a.s.} \infty.
\]
Since $|h_{a,w}|$ is almost surely finite, it follows that
$$\mathbb{P}(\exists N_1:\forall N\ge N_1,\ U_N)=1.$$

\emph{Lower bound.}
We next show that
\[
\min_{\boldsymbol{\phi}} \eta(\boldsymbol{\phi};N)
\xrightarrow[]{a.s.} 0 .
\]
This is established in Lemma~\ref{lemma:minGoToZero} below.
Since $|h_{a,w}|$ is a continuous, nonnegative random variable,
$$\mathbb{P}(\exists N_2:\forall N\ge N_2,\ L_N)=1.$$

\textcolor{black}{
Combining the upper and lower bounds gives
\begin{equation*}
\mathbb{P}\left(\exists N_0:\forall N\ge N_0,\ S_N\right)=1,
\end{equation*}
which proves the theorem.
}
\end{IEEEproof}

\begin{lemma}\label{lemma:minGoToZero}
Let $\{z_i\}_{i=1}^{N}$ be i.i.d. complex random variables defined as
$z_i=g_{s,w,i}h_{a,s,i}$, with $\mathbb{E}[|z_i|]=\mu_z>0$ and
$\mathrm{Var}(|z_i|)<\infty$.
Then
\[
\min_{\boldsymbol{\phi}\in[0,2\pi)^N}
\left|\sum_{i=1}^{N} z_i e^{j\phi_i}\right|
\xrightarrow[]{a.s.}0
\quad \text{as } N\to\infty.
\]
\end{lemma}
 \begin{IEEEproof}
{\color{black}
Fix a realization and set $a_i\triangleq |z_i|$, $A_N\triangleq\sum_{i=1}^{N}a_i$, and $M_N\triangleq\max_{1\le i\le N}a_i$.
Since $z_i=a_i e^{j\theta_i}$, we have $z_i e^{j\phi_i}=a_i e^{j(\theta_i+\phi_i)}$.
Because $\phi_i$ is arbitrary, the term $\theta_i+\phi_i$ can be treated as an arbitrary phase.
For any fixed nonnegative lengths $\{a_i\}_{i=1}^{N}$, the polygon inequality gives
\begin{equation*}
\min_{\boldsymbol{\phi}}
\left|
\sum_{i=1}^{N}a_i e^{j\phi_i}
\right|
=
\max\{2M_N-A_N,0\}.
\end{equation*}
Indeed, the reverse triangle inequality gives the lower bound
\begin{equation*}
\left|
\sum_{i=1}^{N}a_i e^{j\phi_i}
\right|
\ge
M_N-(A_N-M_N)
=
2M_N-A_N,
\end{equation*}
and the magnitude is also nonnegative.
If $M_N>A_N-M_N$, the lower bound is attained by orienting the largest complex vector opposite to all remaining complex vectors.
If $M_N\le A_N-M_N$, the lengths $\{a_i\}_{i=1}^{N}$ satisfy the generalized polygon inequality, so they can be arranged as the sides of a closed polygon.
Orienting the corresponding complex vectors along the directed polygon edges then yields $\sum_{i=1}^{N}a_i e^{j\phi_i}=0$.
This proves the displayed identity.

It remains to show that the zero case occurs eventually almost surely.
By the Strong Law of Large Numbers,
\begin{equation*}
\frac{A_N}{N}
\xrightarrow[]{a.s.}
\mu_z>0.
\end{equation*}
Moreover, $M_N/N\xrightarrow[]{a.s.}0$.
To see this, fix $\epsilon>0$.
Since $\mathbb{E}[|z_1|]=\mu_z<\infty$, the tail-sum bound for nonnegative random variables gives~\cite{durrett2019probability}
\begin{equation*}
\sum_{n=1}^{\infty}\mathbb{P}(|z_n|>\epsilon n)
=
\sum_{n=1}^{\infty}\mathbb{P}(|z_1|>\epsilon n)
\le
\frac{\mathbb{E}[|z_1|]}{\epsilon}
<\infty.
\end{equation*}
By the Borel--Cantelli lemma, $|z_n|>\epsilon n$ occurs only finitely often, almost surely.
Hence, for almost every realization, there exists $n_{\epsilon}$ such that $|z_n|\le\epsilon n$ for all $n\ge n_{\epsilon}$.
Let $C_{\epsilon}\triangleq \max_{1\le n<n_{\epsilon}}|z_n|$.
Then, for all $N\ge n_{\epsilon}$,
\begin{equation*}
\frac{M_N}{N}
\le
\max\left\{\frac{C_{\epsilon}}{N},\epsilon\right\}.
\end{equation*}
Taking $N\to\infty$ gives $\limsup_{N\to\infty}M_N/N\le\epsilon$.
Since $\epsilon>0$ is arbitrary, $M_N/N\xrightarrow[]{a.s.}0$.
Hence, almost surely, there exists $N_0$ such that $2M_N\le A_N$ for all $N\ge N_0$.
For all such $N$, the polygon identity gives
\begin{equation*}
\min_{\boldsymbol{\phi}\in[0,2\pi)^N}
\left|
\sum_{i=1}^{N}z_i e^{j\phi_i}
\right|
=0.
\end{equation*}
This proves the lemma.
}
\end{IEEEproof}

\section{Proof of Lipschitz Gradient} \label{appendix:Lipschitz}
\begin{definition}{\cite[Sec.~1.2.2]{nesterov2004introductory}}
A twice continuously differentiable function $f:\mathbb{R}^N \to \mathbb{R}$ has an $L_f$-Lipschitz continuous gradient if $\|\nabla^2 f(\boldsymbol{\phi})\|_2 \le L_f$ for all $\boldsymbol{\phi}$.
\end{definition}

We derive a bound for the Hessian $\mathbf{H}(\boldsymbol{\phi}) \triangleq \nabla^2 |S(\boldsymbol{\phi})|^2$.
\textcolor{black}{
For $i\neq k$, differentiating the $i$th gradient component with respect to $\phi_k$ gives
\begin{equation*}
[\mathbf{H}(\boldsymbol{\phi})]_{ik}
=
2|z_i||z_k| \cos(\phi_i+\theta_i-\phi_k-\theta_k).
\end{equation*}
The diagonal entries are
\begin{equation*}
[\mathbf{H}(\boldsymbol{\phi})]_{ii}
=
-2|z_i|\sum_{m\neq i}|z_m|\cos(\phi_i+\theta_i-\phi_m-\theta_m).
\end{equation*}
Since $|\cos(x)|\le1$, 
\begin{align*}
|[\mathbf{H}(\boldsymbol{\phi})]_{ik}| &\le 2|z_i||z_k| \quad (i \neq k), \\
|[\mathbf{H}(\boldsymbol{\phi})]_{ii}| &\le 2|z_i|\sum_{m \neq i} |z_m|.
\end{align*}
Since $\mathbf{H}(\boldsymbol{\phi})$ is real and symmetric, $\|\mathbf{H}(\boldsymbol{\phi})\|_2\leq \|\mathbf{H}(\boldsymbol{\phi})\|_{\infty}$. 
Thus,
\begin{equation*}
\|\mathbf{H}(\boldsymbol{\phi})\|_2
\leq L_0, \qquad
L_0 \triangleq 4\max_i \left(|z_i|\sum_{m\neq i}|z_m|\right).
\end{equation*}
The constant $L_0$ is finite for every fixed channel realization with finite $\{z_i\}_{i=1}^{N}$ and is independent of $\boldsymbol{\phi}$.
Therefore, $\nabla |S(\boldsymbol{\phi})|^2$ is Lipschitz continuous.}
\qed

The same conclusion holds for the full objective $\textcolor{black}{P_w(\boldsymbol{\phi})}=|S(\boldsymbol{\phi})+h_{a,w}|^2$, since its Hessian equals $\nabla^2|S(\boldsymbol{\phi})|^2$ plus a diagonal matrix with entries $-2\operatorname{Re}\!\left(h_{a,w}^* z_k e^{j\phi_k}\right)$. 
Therefore, $\|\nabla^2 \textcolor{black}{P_w(\boldsymbol{\phi})}\|_2 \le L_w$, where $L_w \triangleq L_0 + 2|h_{a,w}|\max_i |z_i|$, and $\textcolor{black}{\nabla P_w}$ is $L_w$-Lipschitz continuous.

\section{Gradient Evaluation for Algorithm~\ref{Schem-algo}}
\label{appendix:gradient}
In this appendix, we derive the closed-form gradient used in
Algorithm~\ref{Schem-algo}. Recall that the objective is given by
\begin{equation*}
\textcolor{black}{P_w(\boldsymbol{\phi})}
=
\left|
\sum_{i=1}^{N} z_i e^{j\phi_i}+h_{a,w}
\right|^2 .
\end{equation*}
Define the residual effective channel at Willie as
\begin{equation*}
r(\boldsymbol{\phi})
\triangleq
\sum_{i=1}^{N} z_i e^{j\phi_i}+h_{a,w}.
\end{equation*}
Then,
\begin{equation*}
\textcolor{black}{P_w(\boldsymbol{\phi})}
=
r(\boldsymbol{\phi})r^*(\boldsymbol{\phi}).
\end{equation*}

We now compute the derivative with respect to a single phase variable
$\phi_k$. Since only the $k$th reflected component depends on
$\phi_k$, we have
\begin{equation*}
\frac{\partial r(\boldsymbol{\phi})}{\partial \phi_k}
=
jz_k e^{j\phi_k},
\end{equation*}
and
\begin{equation*}
\frac{\partial r^*(\boldsymbol{\phi})}{\partial \phi_k}
=
-jz_k^* e^{-j\phi_k}.
\end{equation*}
Using the product rule, we obtain
\begin{align*}
\frac{\partial \textcolor{black}{P_w(\boldsymbol{\phi})}}{\partial \phi_k}
&=
\frac{\partial r(\boldsymbol{\phi})}{\partial \phi_k}
r^*(\boldsymbol{\phi})
+
r(\boldsymbol{\phi})
\frac{\partial r^*(\boldsymbol{\phi})}{\partial \phi_k} \nonumber \\
&=
jz_k e^{j\phi_k}r^*(\boldsymbol{\phi})
-
jr(\boldsymbol{\phi})z_k^* e^{-j\phi_k}.
\end{align*}

Let
\begin{equation*}
a_k(\boldsymbol{\phi})
\triangleq
z_k e^{j\phi_k}r^*(\boldsymbol{\phi}).
\end{equation*}
Then,
\begin{equation*}
a_k^*(\boldsymbol{\phi})
=
r(\boldsymbol{\phi})z_k^* e^{-j\phi_k}.
\end{equation*}
Therefore,
\begin{align*}
\frac{\partial \textcolor{black}{P_w(\boldsymbol{\phi})}}{\partial \phi_k}
&=
ja_k(\boldsymbol{\phi})-ja_k^*(\boldsymbol{\phi}) \nonumber \\
&=
j\left(a_k(\boldsymbol{\phi})-a_k^*(\boldsymbol{\phi})\right).
\end{align*}
Since
\begin{equation*}
a_k(\boldsymbol{\phi})-a_k^*(\boldsymbol{\phi})
=
2j\operatorname{Im}\left\{a_k(\boldsymbol{\phi})\right\},
\end{equation*}
we obtain
\begin{equation*}
\frac{\partial \textcolor{black}{P_w(\boldsymbol{\phi})}}{\partial \phi_k}
=
-2\operatorname{Im}
\left\{
a_k(\boldsymbol{\phi})
\right\}.
\end{equation*}
Substituting the definition of $a_k(\boldsymbol{\phi})$ gives
\begin{equation*}
\frac{\partial \textcolor{black}{P_w(\boldsymbol{\phi})}}{\partial \phi_k}
=
-2\operatorname{Im}
\left\{
z_k e^{j\phi_k}r^*(\boldsymbol{\phi})
\right\}.
\end{equation*}
Equivalently, for any $k=1,\ldots,N$,
\begin{equation*}
\left[\nabla \textcolor{black}{P_w(\boldsymbol{\phi})}\right]_k
=
-2\operatorname{Im}
\left\{
z_k e^{j\phi_k}
\left(
\sum_{i=1}^{N} z_i e^{j\phi_i}+h_{a,w}
\right)^*
\right\}.
\end{equation*}

This expression also explains the linear complexity of the gradient
evaluation. At iteration $i$, the residual
\begin{equation*}
r^{(i)}
=
\sum_{k=1}^{N} z_k e^{j\phi_k^{(i)}}+h_{a,w}
\end{equation*}
is computed once. Then all gradient entries are obtained as
\begin{equation*}
\left[\nabla \textcolor{black}{P_w(\boldsymbol{\phi}^{(i)})}\right]_k
=
-2\operatorname{Im}
\left\{
z_k e^{j\phi_k^{(i)}}\left(r^{(i)}\right)^*
\right\},
\quad k=1,\ldots,N.
\end{equation*}
Thus, computing the residual requires $\mathcal{O}(N)$ operations, and
evaluating all gradient entries requires another $\mathcal{O}(N)$
operations. Hence, each gradient-descent iteration has computational
complexity $\mathcal{O}(N)$.

\section{Proof of Lemma \ref{lemma:saddlemax}}\label{Appendix:Lemma8}
\begin{IEEEproof}
Let $f(\boldsymbol{\phi}) \triangleq |S(\boldsymbol{\phi})|^2$, and let $\boldsymbol{\phi}^*$ be a suboptimal critical point with $f(\boldsymbol{\phi}^*) > 0$. By global phase invariance, assume w.l.o.g. that $S(\boldsymbol{\phi}^*) = R > 0$. 
By Lemma~\ref{lemma:critical aligned} (collinearity), each term is aligned with the real axis: $z_k e^{j\phi_k^*} = s_k a_k$, where $a_k \triangleq |z_k|$ and $s_k \in \{+1, -1\}$. Let $\mathcal{P} \triangleq \{k : s_k = +1\}$.

\emph{Step 1: $|\mathcal{P}| \ge 2$.}
Assume for contradiction that $\mathcal{P} = \{p\}$. Then $R = a_p - \sum_{k \neq p} a_k > 0$. For any $\boldsymbol{\phi}$, the triangle inequality yields:
\[
|S(\boldsymbol{\phi})| \ge \left| |z_p| - \left| \sum_{k \neq p} z_k e^{j\phi_k} \right| \right| \ge a_p - \sum_{k \neq p} a_k = R.
\]
This implies $\boldsymbol{\phi}^*$ is a global minimizer, contradicting the premise. Thus, $|\mathcal{P}| \ge 2$.

\emph{Step 2: Negative curvature.}
Select distinct $p, q \in \mathcal{P}$ and define $\mathbf{u} \in \mathbb{R}^N$ such that $u_p = a_q$, $u_q = -a_p$, and $u_k = 0$ otherwise. 
Let $\mathbf{H} = \nabla^2 f(\boldsymbol{\phi}^*)$. Using Appendix~\ref{appendix:Lipschitz}, since $p, q \in \mathcal{P}$ are phase-aligned ($\cos(\phi_p^* - \phi_q^*) = 1$), the relevant Hessian entries are $H_{pq} = 2a_p a_q$ and $H_{ii} = -2a_i(R - a_i)$ for $i \in \{p, q\}$. 

Evaluating the quadratic form for $\mathbf{u}$ yields:
\begin{align*}
\mathbf{u}^T \mathbf{H} \mathbf{u} 
&= H_{pp} u_p^2 + 2H_{pq} u_p u_q + H_{qq} u_q^2 \\
&= -2a_p(R - a_p)a_q^2 - 4a_p^2 a_q^2 - 2a_q(R - a_q)a_p^2 \\
&= -2R a_p a_q (a_p + a_q).
\end{align*}
Since $R > 0$ and $a_p, a_q > 0$, it strictly follows that $\mathbf{u}^T \mathbf{H} \mathbf{u} < 0$. By the Rayleigh quotient, $\lambda_{\min}(\mathbf{H}) \le \mathbf{u}^T \mathbf{H} \mathbf{u} / \|\mathbf{u}\|_2^2 < 0$, establishing that $\boldsymbol{\phi}^*$ is a strict saddle.
\end{IEEEproof}


\bibliographystyle{IEEEtran}
\bibliography{References}
\end{document}